\documentclass[acmsmall]{acmart}

\AtBeginDocument{%
  \providecommand\BibTeX{{%
    \normalfont B\kern-0.5em{\scshape i\kern-0.25em b}\kern-0.8em\TeX}}}

\setcopyright{acmlicensed}
\copyrightyear{2024}
\acmYear{2024}
\acmDOI{XXXXXXX.XXXXXXX}

\acmJournal{JACM}
\acmVolume{1}
\acmNumber{1}
\acmArticle{1}
\acmMonth{1}

\usepackage{hyperref}
\usepackage{footmisc}
\usepackage[T1]{fontenc}% optional T1 font encoding
\usepackage{epstopdf}
\usepackage{graphicx}
\usepackage{placeins}
\usepackage{float}
\usepackage{nomencl}
\usepackage{color,soul}
\usepackage{multirow}
\usepackage{makecell}
\hypersetup{pdfauthor=author}
\usepackage{longtable}
\usepackage{hhline}
\usepackage{pifont}
\usepackage{graphics}
\usepackage{soul}
\usepackage{tabularx}
\usepackage[inline]{enumitem}
\usepackage{geometry}
\usepackage{xurl}
\usepackage{xspace}
\usepackage[binary-units,per-mode=symbol,exponent-product=\cdot,list-units=single,list-final-separator={, }]{siunitx}
\DeclareSIUnit[number-unit-product = ]\pixel{p}
\usepackage{chemformula}
\usepackage{gensymb}
\usepackage{amsmath}
\usepackage[inline]{enumitem}
\usepackage{booktabs}
\usepackage{comment}
\usepackage{footnote}
\usepackage{tablefootnote}
\usepackage{graphicx}
\usepackage{caption}
\usepackage{wrapfig, lipsum}

\expandafter\newcommand\csname r@tocindent4\endcsname{4in}
\newcommand{\ie}{\emph{i.e.}, }
\newcommand{\eg}{\emph{e.g.}, }

\newcommand{\etal}{\emph{et~al.}\xspace}
\begin{document}

\title{A Survey on Energy Consumption and Environmental Impact of Video Streaming}

\author{Samira Afzal$^{\dagger}$, Narges Mehran$^{\dagger}$, Zoha Azimi Ourimi$^{\dagger}$, Farzad Tashtarian$^{\star}$, Hadi Amirpour$^{\star}$, Radu Prodan$^{\dagger}$, Christian Timmerer$^{\star }$}
    \affiliation{ 
      \institution{$^{\star}$Christian Doppler Laboratory ATHENA, Alpen-Adria-Universität Klagenfurt  \country{Austria} }}

\affiliation{%
  \institution{$^{\dagger}$Institute of Information Technology (ITEC), Alpen-Adria-Universität Klagenfurt}
  \streetaddress{}
  \city{Klagenfurt}
  \country{Austria}}
\email{samira.afzal@aau.at}

\renewcommand{\shortauthors}{Afzal, et al.}
\setlength{\parskip}{-1.5pt}
\setlength{\itemsep}{-1pt}
% \setlength{\quotesep}{0pt}
%\titlespacing{\section}{0pt}{0.2\baselineskip}{0.2\baselineskip}
%\titlespacing{\subsection}{0pt}{0.1\baselineskip}{0.1\baselineskip}
%\titlespacing{\subsubsection}{0pt}{0.05\baselineskip}{0.05\baselineskip}
%\titlespacing{\paragraph}{0pt}{0.1\baselineskip}{0.1\baselineskip}

\begin{abstract}
Climate change challenges require a notable decrease in worldwide greenhouse gas (GHG) emissions across technology sectors. Digital technologies, especially video streaming, accounting for most Internet traffic, make no exception. Video streaming demand increases with remote working, multimedia communication services (\eg WhatsApp, Skype), video streaming content (\eg YouTube, Netflix), video resolution (4K/8K, 50\,fps/60\,fps), and multi-view video, making energy consumption and environmental footprint critical. This survey contributes to a better understanding of sustainable and efficient video streaming technologies by providing insights into the state-of-the-art and potential future directions for researchers, developers and engineers, service providers, hosting platforms, and consumers. We widen this survey's focus on \emph{content provisioning} and \emph{content consumption} based on the observation that continuously active network equipment underneath video streaming consumes substantial energy independent of the transmitted data type. We propose a taxonomy of factors that affect the energy consumption in video streaming, such as encoding schemes, resource requirements, storage, content retrieval, decoding, and display. We identify notable weaknesses in video streaming that require further research for improved energy efficiency:
\begin{enumerate*}
\item fixed bitrate ladders in HTTP live streaming; 
\item inefficient hardware utilization of existing video players;
\item lack of comprehensive open energy measurement dataset covering various device types and coding parameters for reproducible research. 
\end{enumerate*}

\end{abstract}




\maketitle

\section{Introduction}
{\label{sec:introduction}}

\emph{Global greenhouse gas (GHG)} emissions impose significant climate change and environmental warming, severely affecting ecosystems and human well-being. Internet data traffic alone accounts for about \qty{3.7}{\percent} of GHG, comparable to the global airline industry~\cite{griffiths2020internet}. One main driver of data traffic is video streaming, responsible for more than \qty{65}{\percent} of the total data volume on the Internet~\cite{Sandvine_Jan2023}. %80\% of the global bandwidth~\cite{2020Globalreport}. %Latency-sensitive and bandwidth-intensive video streaming services are among today's dominating high-velocity traffic generators. 

\begin{figure}[t]
	\centering
	\includegraphics[width=\textwidth]{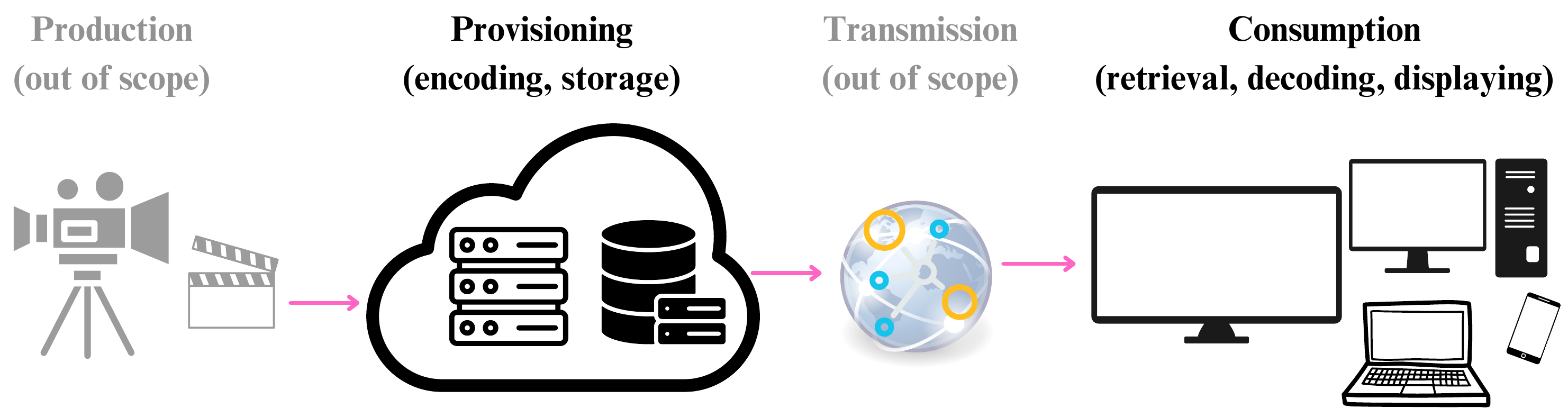}
	\caption{Surveyed video streaming components from production to end-user consumption, comprising encoding, storage, retrieval, decoding, and display.}
	\label{fig:videoProcess}
\end{figure}

\emph{Video streaming} encompasses various applications, including video-on-demand~(VoD) and live services, \qty{360}{\degree} videos~\cite{fan2019survey},  user-generated content platforms, social media platforms, video chats, online meetings, and online games. The demand for video streaming has increased dramatically, mainly since the COVID-19 pandemic, as people increasingly rely on digital online tools for work, education, and entertainment. Therefore, exploring the GHG emissions and energy consumption associated with the production and consumption of video streaming is necessary to mitigate its environmental impact. The video streaming process includes several phases with different energy requirements and environmental impacts, abstractly represented in Figure~\ref{fig:videoProcess}.

\begin{description}[font=\normalfont\itshape]
    \item[Content production] involves filmmaking, editing, and preparation of original video content for distribution~\cite{lopera2021green}; %Energy consumption in this phase depends on the type and scale of the content, the equipment and software used, and the location and duration of the filming. 
    % https://origostudios.com/green-filming/
    % https://origostudios.com/docs/origo-green-executive-report_202108.pdf
    % https://greenfilmshooting.net/blog/en/2015/03/30/energy-efficiency-in-the-italian-film-industry/
    \item[Content provisioning] involves encoding, packaging, and storing videos in different formats, resolutions, and bitrates and storing  them on a Content Delivery Network (CDN); 
    \item[Content transmission] involves delivering video across cloud and edge data centers and to end-user devices through the core Internet, CDNs, and access networks; 
    \item \textit{Content consumption} involves decoding and displaying video content received on end devices. %These devices include laptops, tablets, smartphones, and TVs, with different hardware specifications (such as CPU and GPU capabilities) and software types~(including players and operating systems).   
\end{description}

The current state of digital infrastructure reflects notable progress in energy efficiency and resource utilization. For example, data center servers have become more energy efficient thanks to their cooling system and network equipment~\cite{naffziger2016energy}. Furthermore, despite more demand for Internet service provider (ISP) networks with decreasing relative energy
consumption~\cite{gsma2023mobile}, their electricity consumption is still significant. For example, according to the International Energy Agency (IEA)~\cite{IEA2020}, data centers and networks account for approximately \qtyrange{300}{500}{\mega\tonne\ch{CO2}e} (megatonnes of \ch{CO2} equivalent emissions) in \num{2020}, equivalent to \qty{0.9}{\percent} of energy-related GHG emissions (or \qty{0.6}{\percent} of total GHG emissions). Notably, this consumption corresponds to \qtyrange{2}{3}{\percent} of global electricity use, with data centers and data transmission networks contributing \qtyrange{1}{1.5}{\percent} each~\cite{savazzi2022energy}. Moreover, although end-user devices become increasingly energy-efficient due to rapid hardware advancements,  factors such as energy consumption during production, shorter lifespans, mass production, widespread global use, and high-resolution displays undermine their positive impact. As highlighted in IEA research~\cite{IEA2020}, end-user devices account for the majority of energy consumption (\qty{72}{\percent}), followed by data transmission (\qty{23}{\percent}) and data centers (\qty{5}{\percent}).

Not only the data volume (generation, processing, transmission, and consumption) but also the source of electricity that powers the streaming process impacts the environment. Electricity generation is a significant contributor, accounting for approximately \qty{25}{\percent} of total GHG emissions~\cite{tranberg2019real}. However, not all electricity sources have the same carbon footprint, as some rely on non-renewable and polluting natural powerplants~\cite{verma2019biomass} or use renewable and low-carbon energy powerhouses, such as geothermal, biomass, wind, solar, and hydropower~\cite{gmsys2023afzal}. Therefore, shifting from fossil fuels to renewable energy sources at all video streaming stages can decrease its environmental impact. Leading technology companies, such as Google~\cite{googlsustain}, Microsoft~\cite{microsoftsustain}, and Amazon~\cite{awssustain}, have already invested in sustainable data centers that use renewable energy sources and aim to operate with carbon-free energy by 2030~\cite{google-terawatt}. However, this transition may not progress quickly enough to meet the goals of the Dubai Climate Agreement~\cite{cop28}, limiting global warming to well below \qty{1.5}{\celsius}. 
%The European Union (EU) and its member states have committed to a binding goal of achieving a net domestic reduction of at least \SI{55}{\percent} in GHG emissions by \num{2030} compared to \num{1990} levels~\cite{council2020paris}. 
The United Nations (UN) committed at the COP28 UN climate conference in December \num{2023} to a binding goal of tripling renewables capacity and doubling energy efficiency by \num{2030}~\cite{cop28}. %The latest data collected by the Global Carbon Atlas in 2022 shows China is the world's biggest carbon dioxide emitter, producing 11,397 million tonnes. The US was the second biggest, releasing 5,057 million tonnes into the atmosphere. 

Subjective research studies show the willingness of users to sacrifice video quality to save energy~\cite{gnanasekaran2021digital, hossfeld2023greener, lopez2018prediction}.  Hossfeld~\etal~\cite{hossfeld2023greener} introduced the concept of a ``green user'' through a subjective test characterized by the awareness of energy consumption when evaluating video quality. The study shows that green users are satisfied with half the bitrate compared to non-green users.
% for the same amount of allocated resources (\eg bitrate during video streaming), green users consistently rate video quality higher than ``non-green'' users. This willingness to compromise on Quality of Experience (QoE) to slightly reduce their environmental footprint leads green users to achieve the maximum QoE at \SI{7.25}{\mega\bit\per\second}. On the other hand, non-green users attain the maximum mean opinion score at \SI{14.5}{\mega\bit\per\second}, expecting the highest possible quality given the available resources.
%\paragraph*{Designing energy efficient player}
Some efforts enable an \emph{economical mode (eco-mode)} for video players by adding an energy-saving button. For example, France Télévison\footnote{\url{https://www.france.tv/}, last access: Dec. 2023.} implemented an eco-mode in its player at the cost of quality reduction. Similarly, Bitmovin\footnote{\url{https://bitmovin.com/press-room/bitmovin-launches-eco-mode/}, last access: Dec. 2023.} added an eco-mode to its player to power a more sustainable video streaming landscape. Seelinger~\etal~\cite{Seeliger23End-to-End} implemented a green streaming mode on the video \texttt{dash.js} player with a pop-up that approximates the current \ch{CO2} equivalence emission offering to lower the video quality to an acceptable level.

Considering the importance of energy consumption and \ch{CO2} emissions associated with video streaming, we conduct a comprehensive survey of the most valuable studies in two phases:
\begin{enumerate}
\item \emph{Content provisioning} considering video encoding and storage;
\item \emph{Content consumption} exploring video retrieval, decoding, and display.
 \end{enumerate}
Video production remains outside the scope of this survey due to limited available energy-saving solutions~\cite{lopera2021green, starosielski2016sustainable, evans2020sustainable}. Similarly, we exclude transmission since the type of data (\ie video in our case) does not impact the energy consumption of the network data transmission according to existing surveys available on energy consumption in networks~\cite{feng2012survey, ehsan2011survey}.
% \begin{figure}[t]
% 	\centering
% 	\includegraphics[width=\columnwidth]{sections/Figures/methodologySurvey.png}
% 	\caption{Literature selection methodology inspired by~\cite{turner2010does}.}
% 	\label{fig:metotholody}
% \end{figure}
% \subsection{Related work}
% Table~\ref{tab:relatedsurveys} covers the convergence and comparison of the previously published surveys and this survey.
% \begin{table}[!t]
% \caption{Comparison with the previously published surveys.}
% \label{tab:relatedsurveys}
% %\resizebox{\columnwidth}{!}{
% \begin{tabular}{|@{ }l@{ }|@{ }l@{ }|}
% \hline
% &\\
% \hline
% &\\
%https://ieeexplore.ieee.org/abstract/document/1510627
%https://www.sciencedirect.com/science/article/abs/pii/S0065245821000280
%https://ieeexplore.ieee.org/stamp/stamp.jsp?tp=&arnumber=10213996
%https://www.sciencedirect.com/science/article/pii/S092523122030494X?casa_token=UOSsUdAPUysAAAAA:Xj7LP5pU2RJz0F7O2kLXOF4FsD39wFIars6Y9iBiNKu_9X8tAU0TfDEnSN0ft5BPNcHBKa_MCQwO
% \hline
% \end{tabular}
% %}
% \end{table}
\subsection{Literature Selection Methodology}

\begin{figure}[t]
	\centering
	\includegraphics[width=\textwidth]{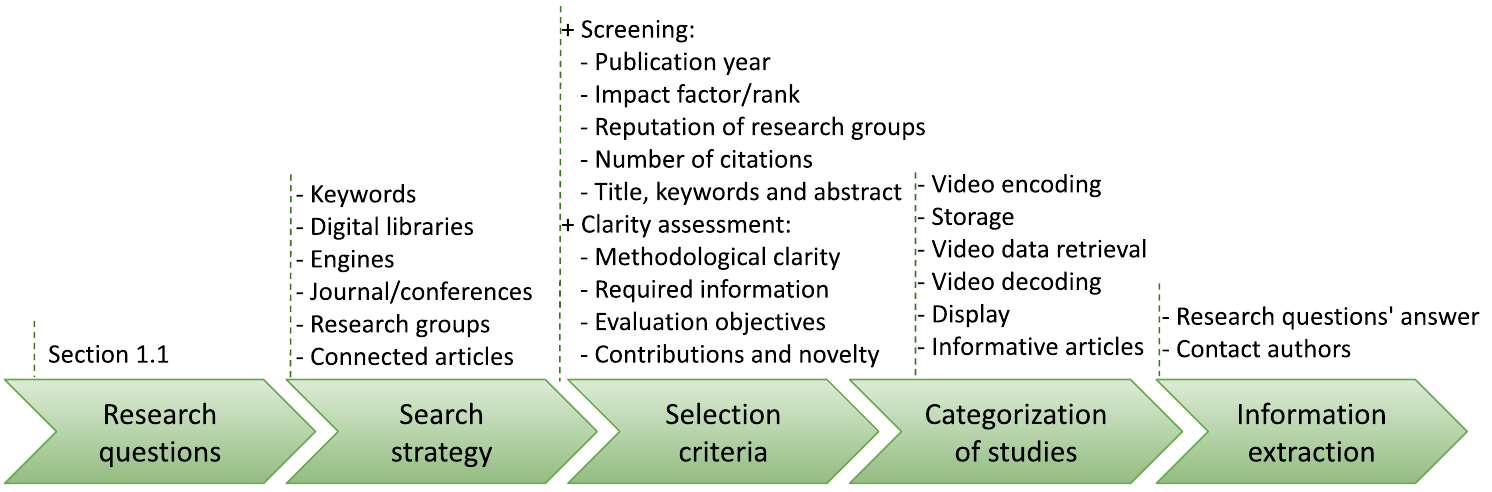}
	\caption{Literature selection methodology inspired from~\cite{turner2010does}.}
	\label{fig:metotholody}
\end{figure}

Figure~\ref{fig:metotholody} describes the strategy for a systematic literature review in this survey inspired by~\cite{turner2010does}.

\paragraph*{Research questions}
The main research questions of this survey are:

\begin{enumerate}[leftmargin=*,align=left]
    \item What is the impact of video streaming on overall \ch{CO2} emissions?

    \item How do various video streaming components contribute to its energy use and \ch{CO2} emissions?
    
    \item What are the main factors that affect the energy consumption of each
    streaming component?

     \item How can we estimate or measure the energy consumption and \ch{CO2} emissions?
     
    \item What are the existing energy or \ch{CO2} estimation models for each video streaming component, and their advantages, disadvantages, accuracy, and reliability?
    
    \item Which tools and datasets can measure the power, energy, and \ch{CO2} emissions of video streaming? 
    
    \item What are the challenges and future opportunities to advance research on energy consumption and \ch{CO2} emissions of video streaming systems? %and what are the potential directions and solutions?
\end{enumerate}

\begin{table}[t]
\small
\caption{Search strategy and sources utilized in the systematic literature review.}
\label{table:methodology}
%\vspace*{-0.3cm}
\resizebox{\textwidth}{!}{
\begin{tabular}{|c|p{12cm}|p{1cm}|}
\hline
\emph{Search}   & \emph{Description} & \emph{Articles} \\ \hline\hline
\multirow{1}{*}{\emph{Video keyword}}  & Streaming; Delivery; Transmission; Encoding (hardware and software); Decoding; Complexity &  \multirow{3}{*}{ {899} }\\
\cline{1-2}
\multirow{2}{*}{\emph{Energy keyword}} & Consumption; Power; \ch{CO2}; Carbon footprint; Carbon emission; Utilization; Efficiency; Waste; Sustainability; Renewable; Green; Awareness; Saving; Measurements; Climate & \\ \hline
\emph{Digital library}   & \multirow{1}{*}{ACM, IEEE, Elsevier}  & \multirow{1}{*}{430} \\ \hline
\emph{Engine}   & Google Scholar, Google  & 620 \\ \hline
\multirow{8}{*}{\makecell{\emph{Multimedia}\\\emph{venue}}}   &
Multimedia Tools and Applications & \multirow{8}{*}{25}\\
& IEEE Transactions on Multimedia &\\
& ACM International Conference on Multimedia &\\
& ACM Transactions on Multimedia Computing, Communications, and Applications &\\
& IEEE International Conference on Multimedia and Expo &\\
& ACM Multimedia Systems Conference &\\
& IEEE Transactions on Circuits and Systems for Video Technology &
\\ \hline
\multirow{3}{*}{\emph{Energy venue}}  & 
Renewable and Sustainable Energy Reviews & \multirow{3}{*}{6}\\
& Energy Strategy Reviews &\\
& Energy Policy &\\
\hline
\emph{Research group}  & Green Streaming~\cite{green-stream-proj}; DIMPACT~\cite{dimpact}; Greening of Streaming~\cite{greening-of-streaming-proj} & 5  \\ \hline
\end{tabular}}
\end{table}

\paragraph*{Search strategy} Table~\ref{table:methodology} summarizes our search strategy, leading to a total of \num{889} articles, including journals and conference papers selected for this survey.

\paragraph*{Keywords} We started by defining relevant keywords, including synonyms and variations, related to major research areas of this article: \emph{multimedia} and \emph{energy}.

\paragraph*{Digital libraries} Our next step was to identify the studies from the pivotal multimedia- or energy-related digital libraries, journals, and proceedings from conferences. We expanded our search beyond these sources, utilizing Google Scholar to uncover articles that might not have been in our initially chosen libraries, considering that some might have appeared in multiple databases.

\paragraph*{Search engine} We did not confine our search to academic sources but also explored Google for video and energy-related project websites, technical reports, and online sources related to our topic, providing additional technical information for our survey.

\paragraph*{Research} Additionally, we examined the works of research groups and individual contributors actively shaping the field to find the latest and most significant articles. We also followed the cross-references and employed a connected articles search approach by examining the references and citations of pre-selected articles.

\paragraph*{Selection strategy}
We defined two steps for selecting the articles. 

\begin{description}[font=\normalfont\itshape]
    \item[Screening.] We applied the following criteria to select articles aligned with the survey's objectives by filtering over \qty{60}{\percent} for quality assurance and keeping the ones: 
    \begin{enumerate*}
        \item recently published (after \num{2019}), driven by the recognition of a substantial and accelerating trend in the research landscape,
        \item with comparable impact factors or ranks of the venues,
        \item authored by a reputable affiliated university or company,
        %(\eg Fraunhofer FOKUS, Ateme, Synamedia),
        \item with over five citations per year. %, and
        %\item the reviewed title, keywords, and abstract.
    \end{enumerate*}
    \item[Clarity assessment.] We assessed the articles by examining methodological and conceptual clarity, transparency in presenting results, inclusion of required information, evaluation objectives, contributions to the field, and novelty. Following this assessment, we narrowed the selection by another \qty{15}{\percent}.
\end{description}

\paragraph*{Categorization}
We divided the collected articles based on their contribution toward content provisioning and consumption and principal components of \begin{enumerate*} \item video encoding, \item storage, \item video content retrieval, \item video decoding, and \item video display\end{enumerate*}. Additionally, we consider informative articles that provide valuable support for our survey but do not directly fit into any specific category.

% We categorized the main scopes of journals and conferences, publishing the research articles covered in this survey into four categories: \emph{Multimedia}, \emph{Sustainability}, \emph{Computing}, and \emph{Communication}. Then, Figure~\ref{fig:cat} illustrates the categorization of selected articles according to the main scopes of the respective journals or conferences. Each bar represents the number of articles for each component in a specific category, providing insights into the distribution of selected articles across these key thematic areas.

% 9 papers for encoding energy
% 14 computing continuum
% 4 for storage
% 27

% 9 NIC
% 12 decoding
% 7 display
% 28
% \begin{wrapfigure}{R}{0pt}
% 	\centering
% 	\includegraphics[width=0.5\columnwidth]{sections/Figures/cat.pdf}
%     \vspace*{-0.9cm}
% 	\caption{Categorization of selected articles based on journal or conference scope.}
% 	\label{fig:cat}
% \end{wrapfigure}

\begin{figure*}[t]
	\centering
	\includegraphics[width=\textwidth]{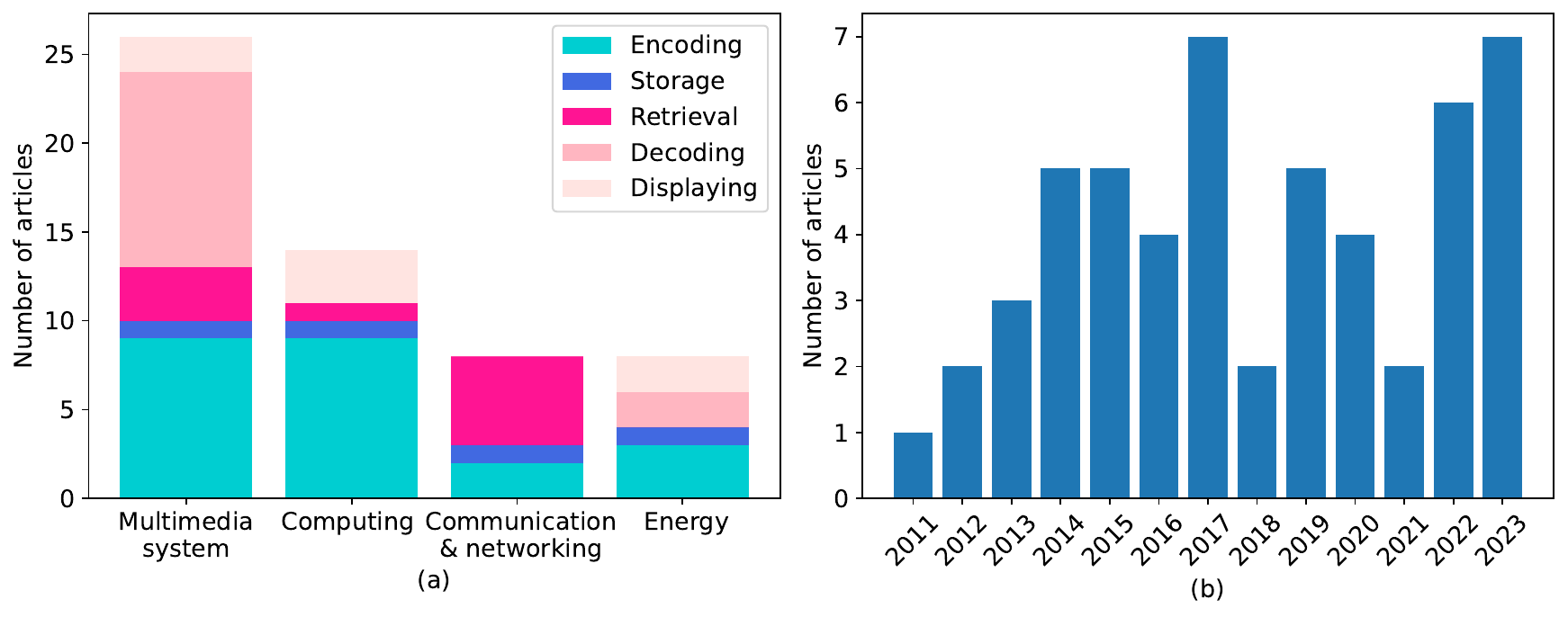}
	\caption{\centering (a) Categorization of articles based on scope, and (b) distribution in the past thirteen years.}
	\label{fig:cat}
\end{figure*}

\paragraph*{Information extraction}
Finally, we extracted data from the selected studies to address each of the research questions described above. We contacted the authors for further detailed information about their proposed methods and experimental designs in case of a requirement.

\paragraph*{Results}
Out of the \num{889} initially selected articles, we thoroughly surveyed \num{56} articles in our work, referring to a total of approximately \num{200} references. We categorized the main scopes of journals and conferences, publishing the research articles covered in this survey into four categories: \emph{multimedia systems}, \emph{computing},  \emph{communication}, and \emph{energy}. Then, we categorized the selected articles %according to the scopes of the respective journals or conferences (see Figure~\ref{fig:cat})
according to the scope of every inspected paper in the respective journal or conference (see Figure~\ref{fig:cat}(a)), illustrating significant research focus and effort directed towards achieving sustainable video streaming within the multimedia community. Moreover, Figure~\ref{fig:cat}(b) depicts the distribution of articles published from \qtyrange{2011}{2023}, highlighting a predominant trend of recent publications. While acknowledging the surge in publications addressing energy considerations in the multimedia field, we have chosen to conclude our review at this point.

\subsection{Contributions}
% Energy consumption in video streaming is a topic that requires a comprehensive survey to evaluate and advance the field. However, there is only one survey paper on this topic~\cite{Hoque2014} published in \num{2014}, which categorized the research works based on the layers of the Internet protocol stack. Therefore, there is a need for a timely and comprehensive assessment of the energy consumption in video streaming.

Although energy consumption in video streaming is a relevant and timely research topic, there is only one survey paper on this topic~\cite{Hoque2014} published in \num{2014}, which categorized the research works based on the layers of the Internet protocol stack. To bridge this gap, this survey covers a comprehensive literature review of energy consumption for video streaming, covering \num{56} research works that offer a comprehensive understanding of the current state of research and open issues. A collection of tools and datasets that target the power, energy, and \ch{CO2} emissions measurements for video researchers and engineers to examine. 
    %  \item A comprehensive study of the challenges in \ch{CO2} emission measurement and energy consumption calculation for video streaming;
    % \item A comprehensive literature review of energy consumption for video streaming covering relevant approaches and new techniques in the field (close to fifty research works), classifying them accordingly based on the computational aspects in the Cloud and edge computing continuum, storage in CDNs,  data receive through the network interface cards (NICs), decoding and display on end-user devices;
    % \item In-depth discussion of the challenges and opportunities to optimize energy consumption in different components of the video streaming process;   
    % \item Introduction of the tools and datasets regarding the power, energy, and \ch{CO2} emissions measurements;
    % \item Outline of several open issues and trends for future research.
%\end{itemize}

\begin{figure}[!t]
	\centering
	\includegraphics[width=\textwidth]{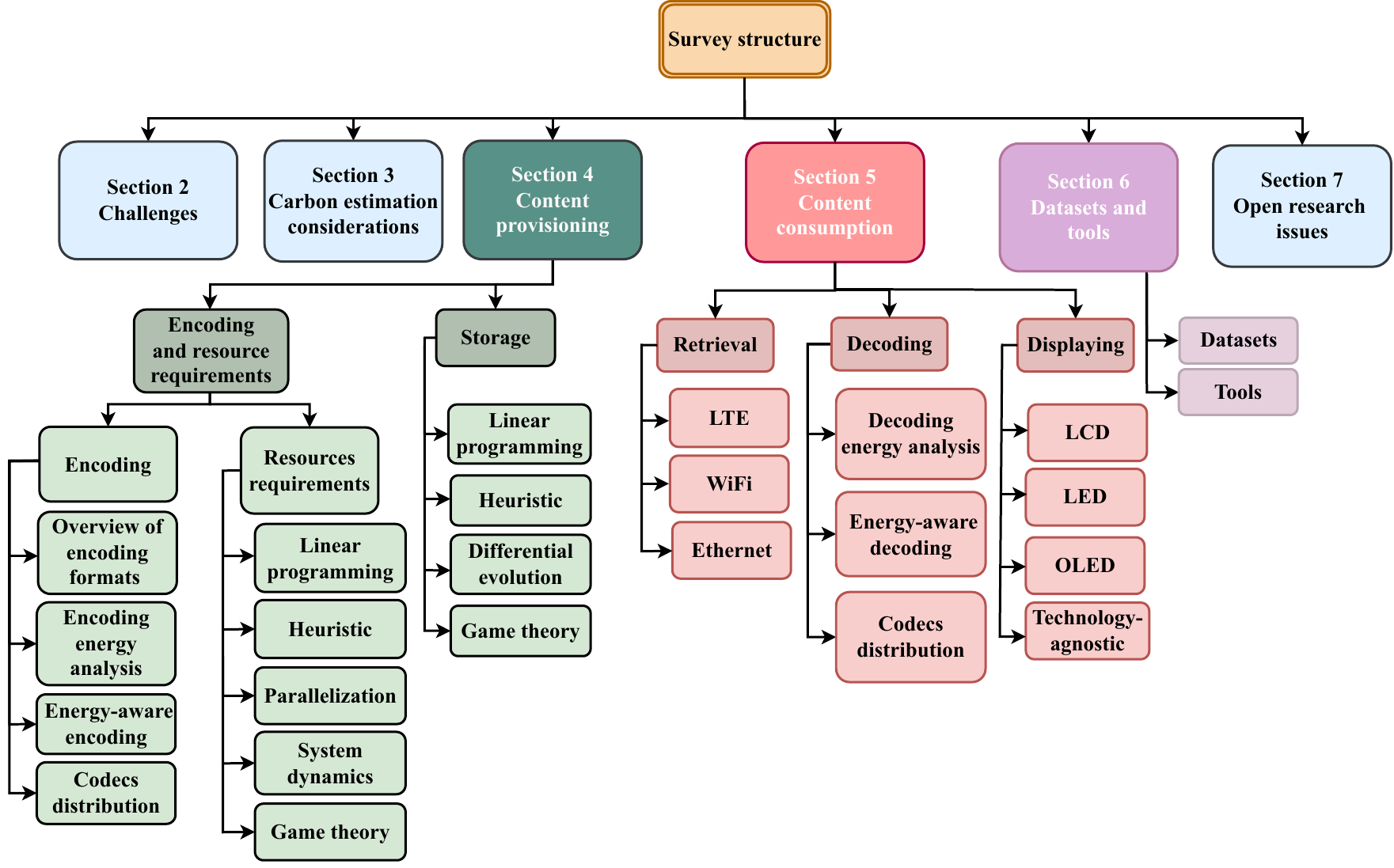}
	\caption{Energy consumption and environmental impact of video streaming survey structure.}
	\label{fig:outline}
\end{figure}

\subsection{Survey structure} Figure~\ref{fig:outline} depicts 
the structure of this survey. Section~\ref{sec:Challenges} presents the challenges of measuring and reducing the carbon footprint and energy consumption of video streaming. Section~\ref{sec:CarbonEstimation} explores the considerations and methods for estimating the carbon emissions of video streaming. Section~\ref{sec:cp} investigates the energy consumption of the components of the content provisioning phase involved in the encoding and storage of video. Section~\ref{sec:cc} studies the energy consumption of the components of the content consumption phase involved in receiving, decoding, and displaying video content. Section~\ref{sec:dataset&tools} provides the available datasets and tools for measuring energy consumption and carbon emissions. Section~\ref{sec:openResearches} discusses research issues and directions for energy-efficient and sustainable video streaming. Finally, Section~\ref{sec:Conc} provides concluding remarks.%A list of abbreviations can be found in the appendix to help readers handle the myriad of acronyms.

\section{Challenges}{\label{sec:Challenges}}  
Measuring and reducing the energy consumption and carbon footprint of video streaming is not trivial, as many factors and uncertainties affect it~\cite{afzalChall2023,IEA2020}.  This section discusses the main challenges researchers face in understanding the energy and \ch{CO2} emissions associated with video streaming.

\subsection{Variability in carbon footprint measurements} The carbon footprint of video streaming depends on several factors that can vary widely, such as location, device, technology, and network energy intensity~\cite{trust2021}:

\paragraph*{Country-specific electricity emission} is the factor with the most significant variability of the electricity grid, reflecting carbon produced per unit of electricity generated. For example, Germany's grid emission factor is 30 times higher than Sweden's, leading to a 30-time difference in the overall carbon footprint~\cite{trust2021}. The carbon intensity of electricity also changes over time as the electricity supply becomes more or less dependent on fossil fuels.

\paragraph*{End-user device} is the second most influential variability factor. For example, the carbon footprint of a 50-inch TV is about 4.5 higher than a laptop and 90 times higher than a smartphone~\cite{trust2021}.

\paragraph*{Estimation year} is significant, as technological advances and the decarbonization of electricity grids lead to a decrease in energy intensity and emission factors. 
 
\paragraph*{Network energy intensity} factors may vary by operator, country, the age of the network equipment, population density, and even climatic factors (\ie surroundings temperature and humidity).

\subsection{Uncertainty for carbon footprint measurements} 

%\begin{wraptable}{R}{.6\columnwidth}%[!t]
%\footnotesize
\begin{table}
\small
\caption{\centering Carbon emission estimation for one hour of video streaming according to geographic scope.}
\label{tab:CarbonEmission}
%\resizebox{.6\textwidth}{!}{
\begin{tabular}{|@{ }l@{ }|@{ }l@{ }|@{ }l@{ }|@{ }l@{ }|@{ }l@{ }|@{ }l@{ }|@{ }l@{ }|}
\hline
\multirow{2}{*}{\emph{Work}} & \multirow{2}{*}{\emph{Year}} & \multirow{2}{*}{\emph{Geo-scope}}  & \multicolumn{3}{l|}{\emph{Scenario}} & {\emph{\ch{CO2} emission }}\\
\cline{4-6}
&& &\emph{Receiver} & \emph{Resolution} & \emph{Network} & [\si{\gram\per\hour}]\\
\hline\hline

    Makonin~\cite{makonin2022calculating} & 2022 & Global& TV&\SI{720}{\pixel}& WiFi& 339 – 425    \\  \hline
    IEA~\cite{kamiya2020carbon}& 2019 & Global& Mix &Mix& Mix& 36\\ \hline

  Marks~\cite{miller2020streaming}& 2020  & Global&  PC &\SI{1080}{\pixel}&-- & 77000   \\ \hline

  Shift Project~\cite{efoui2019climate} & 2018  & Global & Mix &Mix& Mix&394   \\  \hline

  BITKOM~\cite{coroama2020nachhaltigkeit} & 2018 & Global&TV (65") & \SI{2160}{\pixel} &Landline & 610    \\ \hline

  BITKOM~\cite{coroama2020nachhaltigkeit} & 2018  & Global & TV (65") & \SI{720}{\pixel} &Landline & 130   \\ \hline
  \hline
   Carbon Trust~\cite{trust2021} & 2020 & EU  &  Mix  &Mix&Mix &56    \\ \hline

  Carbon Trust~\cite{trust2021} & 2020 & Germany & Mix  &Mix& Mix& 76    \\ \hline
  Carbon Trust~\cite{trust2021} & 2020 & UK &  Mix&Mix&Mix & 48    \\ \hline

  BBC~\cite{fletcher2021carbon}  & 2019--2020 & UK & TV (Player) &--&-- &33   \\ \hline

  BBC~\cite{fletcher2021carbon} & 2019--2020 & UK& TV (IPTV) &--& --&  32    \\ \hline

  BBC~\cite{fletcher2021carbon} & 2019--2020 & UK& TV &\multirow{1}{*}{--}&Terrestrial &  17    \\ \hline

  IEA~\cite{kamiya2020carbon} & 2019 & Germany & Mix&Mix&Mix & 31    \\  \hline

%\hline
\end{tabular}
%}
\end{table}

Uncertainty refers to the degree of precision of the measurements.
Carbon emission estimation studies often exhibit uncertainty due to differences in their allocation methods, the limited amount of publicly available data, and the tools used. This uncertainty in estimations is evident in Table~\ref{tab:CarbonEmission} that summarizes the carbon emission estimations for one hour of video streaming from different regions (\ie Global, EU, Germany and UK) and years published (\numrange{2018}{2022}) in various studies. These estimates vary widely, from \num{36} to a highly uncertain value of \SI{77000}{\gram\per\hour} of \ch{CO2}, depending on their assumptions, methods, choice of devices, network connection, and video resolution.

\subsection{Decoupling data growth from energy consumption} 
Experts argue that increasing the demand for services, like video streaming as a data-intensive activity, does not necessarily increase energy consumption~\cite{streaming-carbon-sfu}, since cloud providers increasingly use renewable electricity, with some reaching \SI{100}{\percent} renewable targets. Similarly, major telecommunication network operators are setting \SI{100}{\percent} renewable targets and \SI{1.5}{\celsius} compatible science-based targets. Technological advances are making it possible to design more energy-efficient end-user devices.
Interestingly, academic studies~\cite{EGG2020,DecarbonisingData2020,Stobbe2021} and network operator reports in \num{2020} showed that data traffic experienced a significant increase with minimal impact on energy use. For example, telecommunication companies reported less than \SI{1}{\percent} increase in energy consumption, despite data traffic growing by \SI{50}{\percent}~\cite{GSMA2020}. Telefonica, the Spanish multinational telecommunications company, reported a \SI{45}{\percent} increase in data traffic due to COVID-19, with a slight decrease in energy consumption~\cite{telefonica2020}. Similarly, Cogent, a large operator of fiber-optic backbone networks in the USA, reported a \SI{38}{\percent} increase in data traffic and a decrease in overall network energy use~\cite{cogent2020}. From these reports, we infer no necessary correlation between data traffic and network energy consumption~\cite{trust2021}.

% Finally, networks and data centers operate continuously regardless of the data usage.

\section{Carbon estimation considerations}
\label{sec:CarbonEstimation}

This section discusses the factors that affect the calculation and comparison of the greenhouse gas emissions associated with video streaming. These factors include 
% the definition and measurement of “gCO2e”, the unit of carbon footprint, and
the variation and optimization of the carbon intensity of electricity generation and consumption depending on the energy sources, the imported electricity, and the timing of electricity use~\cite{google2022reducing}. These considerations are essential for understanding and reducing the environmental impact of video streaming.

\paragraph*{Electricity generation sources} The diverse sources used for electricity generation directly impact each region's carbon intensity. Primary sources include \begin{enumerate*} \item fossil fuels, such as coal, oil, and natural gas, and \item renewable sources, such as solar, wind, hydro, geothermal, and biomass.\end{enumerate*}
Recently, cloud providers are adopting renewable energy sources~\cite{kong2014survey}. For example, Google classified its data centers based on carbon emission characteristics. Moreover, the Amazon Web Services (AWS) region in Oregon and Toronto, \ie \texttt{us-west1} and \texttt{northamerica-northeast2}, are among the \emph{low carbon} regions~\cite{google-carbon-regions}. 
The sustainability report of AWS shows that renewable sources provide power for over \qty{95}{\percent} of six European and seven North American regions in \num{2021}~\cite{awssustain}.
Considering the emissions associated with each source is necessary for calculating the carbon intensity of electricity production. Fossil fuel-based power plants emit significant amounts of \ch{CO2} during combustion. In contrast, renewable energy sources have minimal or zero carbon emissions during operation. Factors such as energy mix, power plant efficiency, and the carbon content of different fuel sources are important to determine the carbon intensity for each region. This kind of data allows for a comprehensive assessment of the carbon footprint associated with electricity generation in specific regions, aiding policy-makers and stakeholders in transitioning to %cleaner and
more sustainable energy.

\paragraph*{Imported electricity} For an accurate calculation of the carbon footprint of electricity imported from other countries, its entire life cycle is essential, from its generation to its consumption. This life cycle requires accounting for the emissions associated with the production of electricity and the emissions associated with its transportation and distribution. For example, most of the UK's imported electricity comes from France via interconnectors~\cite{cheng20225g}.

\paragraph*{Timely electricity use for carbon emission mitigation} Timing electricity use is crucial in leveraging the availability of renewable energy sources. The importance stems from the intermittent nature of certain renewables, such as solar and wind power. Aligning electricity consumption with periods of maximum availability of renewable energy (\eg day sunshine, windy seasons) can significantly reduce the reliance on conventional power sources and maximize the use of %clean and
sustainable alternatives. 
% This concept, called demand response or demand-side management, allows for a more efficient and environmentally friendly grid operation, promoting the integration of renewable energy into our daily lives. Embracing the flexibility of time in electricity usage not only enhances the utilization of renewable resources but also contributes to reducing carbon emissions and fostering a greener energy future.

\section{Content provisioning}
\label{sec:cp}

Content provisioning involves encoding and storage. Video encoding applies compression algorithms to reduce video size and bandwidth while maintaining quality. Reasons for video encoding are the growth of video-related applications such as VoD and live video streaming, the emergence of immersive video technologies, the evolution of video attributes such as spatial resolution and framerate, reducing cost (computation, bandwidth, storage), and improving quality of experience (QoE). Video encoding generally occurs in data centers, where different servers and instances perform the encoding tasks~\cite{netflix2015highquality,bitmovin2015cloud}. The energy consumption of video encoding depends on factors such as codec, resolution, bitrate, framerate, preset, and complexity of the video content. After encoding, storage systems, such as CDN, consume energy to store and access video data.

\subsection{Video encoding  and resource requirements} \label{sec:coding}
This section starts with an overview of different video encoding formats' main features and characteristics. It then presents a detailed analysis of the energy consumption of video encoding processes, considering various factors such as codec, resolution, bitrate, preset, and complexity of the video content. Next, it surveys existing solutions and methods that optimize the energy consumption of video encoding. Finally, it discusses the challenges and opportunities for future research and development in energy-efficient video encoding.

\subsubsection{Video encoding}
\paragraph{Video encoding formats} 
Video encoding is the process of compressing video data using different standards or formats, typically in five stages.
\begin{description}[font=\normalfont\itshape]
\item[Block partitioning] divides the video into smaller blocks for more efficient processing.
\item[Residual prediction] predicts each block either from previously encoded blocks in the same frame (intra-prediction) or from previously encoded blocks in the previous or future reference frames (inter-prediction) to reduce redundancy.
\item[Transformation] converts the residual data into the frequency domain, as the human visual system is less sensitive to high frequencies, allowing for the removal of less important data.
\item[Quantization] reduces the precision of the transformed coefficients to minimize the bitrate.
\item[Entropy coding] compresses the quantized coefficients into the final bitstream.
\end{description}
While all video encoders use these stages to compress video data, the historical evolution of video standards has played a crucial role in shaping the efficiency and quality of video compression. These standards have continually pushed the boundaries of video encoding capabilities, adapting to the increasing demands of modern multimedia.

% https://www.wowza.com/blog/the-7-best-hardware-encoders-for-live-streaming
% https://www.gumlet.com/learn/video-codec/ 
% https://www.linkedin.com/pulse/video-coding-standards-comparison-sraas/
\paragraph*{Advanced Video Coding} AVC~\cite{wiegand_overview_2003}, ratified in \num{2003} revolutionized video compression. AVC supports up to 4k and became the standard for HD television, video streaming, and video conferencing, remaining dominant for a considerable period~\cite{bit-report-2023}.
It introduced advanced motion estimation (ME) and compensation techniques to improve compression efficiency substantially. AVC divides each frame into macroblocks of $16\times 16$ pixels, and to improve motion compensation, a variable block size prediction ($4\times 4$ to $16\times 16$) is used, resulting in improved prediction and compression efficiency. Multi-reference picture motion compensation is also supported in this standard. 
Common AVC software implementations include x264~\cite{x264}, JM~\cite{jm}, and FFmpeg~\cite{ffmpeg}. Popular hardware implementations encompass Intel Quick Sync Video~(QSV)~\cite{IntelQSV}%dedicated hardware core on the processor-INTEL GPU 
,  Nvidia NVENC~\cite{nvidia} (NVDEC for decoding), AMD Video Code Engine (VCE)~\cite{amd-vce}%integrated into all of their GPUs and APU
, and ARM (Media Foundation)~\cite{arm}.

\paragraph*{High-Efficiency Video Coding}
HEVC~\cite{sullivan2012overview} released in \num{2013}, ten years after AVC ratification, offered even greater compression efficiency, well suited for 4K video streaming and UHD Blu-ray.  HEVC compression efficiency increases by approximately \qty{50}{\percent} compared to AVC in cost of \numrange{1.2}{3.2}-fold higher complexity at the encoder side~\cite{vanne2012comparative}.  HEVC processes information in coding tree units (CTUs)~\cite{cetinkaya_ctu_2021}  instead of AVC macroblocks. HEVC CTUs can range from $4 \times 4$ to $64  \times 64$ pixels, allowing for more efficient compression than AVC. Additionally, HEVC features advanced complex motion compensation and improved extra intra-prediction modes compared to AVC~\cite{monteiro_rate-distortion_2015}. 
Typical software implementations of HEVC are x265~\cite{x265}, FFmpeg~\cite{ffmpeg},  HM~\cite{hm} and Kvazaar~\cite{kvazaar}. Common hardware implementations include Intel QSV~\cite{IntelQSV}, Nvidia NVENC~\cite{nvidia} (NVDEC for decoding), AMD VCE~\cite{amd-vce}, and ARM (Media Foundation)~\cite{arm}.

\paragraph*{Versatile Video Coding} VVC~\cite{bross_overview_2021}, introduced in \num{2020}, is another notable addition to the landscape in recent years.  VVC builds upon the achievements of HEVC to handle 8K resolution and immersive video experiences, promising to be a key player in the future of video encoding.  VVC further enhances compression efficiency by reducing the bitrate by \qty{50}{\percent} while maintaining the same visual quality as HEVC. This efficiency improvement comes at the cost of the increased computational complexity of  \qty{50}{\percent}  for both the encoder and decoder~\cite{pakdaman_complexity_2020}. Compared to HEVC, the block size increased from  $64  \times 64$  to $128  \times 128$ pixels, and \num{67} more modes are available.  
Motion compensation in VVC is also improved by introducing new and advanced coding tools. Several software implementations of VVC are VTM~\cite{VVCSoftware}, VVenC~\cite{vvenc} (VVdeC~\cite{vvdec} for decoding). An example hardware implementation is Xilinx Virtex-7 FPGA~\cite{Xilinx}.

\paragraph*{VP9}~\cite{vp9}, released in 2013 by Google, is an open and royalty-free video coding format. Initially, VP9 found primary usage on YouTube and proved to have a \qtyrange{40}{45}{\percent} bitrate advantage over AVC~\cite{vcodex}.  This codec can handle a variety of block sizes, ranging from $6  \times 6$ to $64  \times 64$ pixels. This flexibility allows for more efficient coding of frame parts with different-sized blocks based on the level of detail or motion.
Software implementations of VP9 include Vpxenc~\cite{uitto_energy_2016} and VP9~\cite{vp9}. Hardware implementations differ based on their supports for \begin{enumerate*} \item encoding and decoding (\eg Intel QSV~\cite{IntelQSV}), and \item just decoding (\eg AMD VCE~\cite{amd-vce}, Nvidia NVDEC~\cite{nvidia}, ARM~\cite{armvp9}).\end{enumerate*}

\paragraph*{AV1}~\cite{han_technical_2021} released in 2018 provides efficient video compression while remaining royalty-free. It is gaining adoption in video streaming, especially on platforms prioritizing quality and efficiency. AV1 offers \qty{30}{\percent} better compression than HEVC for the same image quality at the cost of four times longer encoding time. AV1 extends the maximum block size from $64  \times 64$ pixels in VP9 to $128  \times 128$ pixels, allowing for more efficient encoding of high-resolution content.
Software implementations of AV1 include SVT-AV1~\cite{svtav1}.  Hardware implementations are Intel QSV~\cite{IntelQSV}, Nvidia NVENC~\cite{nvidia} (NVDEC for decoding), AMD VCE~\cite{amd-vce}, ARM64~\cite{armvav1}.

\paragraph{Video encoding energy analysis} 
\label{sec:encoding-analysis}
Table~\ref{tbl:encoding} summarizes related works that analyze energy consumption in video encoding. Several factors influence the consumption of video encoding energy, categorized into \begin{enumerate*} \item codec and \item encoding parameters\end{enumerate*}, explained in the following paragraphs.

% standard resolutions: https://en.wikipedia.org/wiki/Display_resolution 
\begin{table*}[!t]
\caption{Related work classification on video encoding (CD: codec, RS: resolution, FR: framerate, PR: preset, QP: quantization parameter).}
\label{tbl:encoding}
\resizebox{\textwidth}{!}{
\begin{tabular}{|l|l|l|l|l|l|l|l|l|l|}
\hline
\multirow{2}{*}{\emph{Work}} & \multirow{2}{*}{\emph{Goal / Method}} & \multicolumn{5}{l|}{\emph{Encoding parameters}} & \multicolumn{2}{l|}{\emph{Machine}} & \emph{Energy}\\
\cline{3-9}
& & \emph{CD} & \emph{RS}& \emph{FR [fps]} & \emph{PR} & \emph{QP} & \emph{Type} & \emph{Processor} & \emph{tool} \\
\hline\hline

\multirow{2}{*}{\cite{silveira_performance_2017}} & {Software energy}  & \multirow{2}{*}{HEVC (x265)} & \multirow{2}{*}{\qty{1080}{\pixel} }
& \multirow{2}{*}{ \numrange[range-phrase=--]{0.4}{22.1}} & \texttt{Veryslow} -- & \multirow{2}{*}{--} &   \multirow{2}{*}{--}& \multirow{2}{*}{Intel CPU} & CACTI ~\cite{cacti2022cacti}  \\
&  analysis& & & & \texttt{ultrafast} & &  & &PCM~\cite{intelpcm} \\

\hline

\multirow{3}{*}{\cite{monteiro_rate-distortion_2015}} & Software 
energy   & AVC (JM) & \qty{416}{}$\times$\qty{240}{}, & 20, 24,&  \multirow{3}{*}{--}& 22, 27, & \multirow{3}{*}{PC} & \multirow{3}{*}{Intel CPU} & \multirow{3}{*}{RAPL~\cite{khan2018rapl}} \\
&analysis & HEVC (HM) & \qty{832}{}$\times$\qty{480}{}, & 30, 50, & & 32, 37 &  & &  \\
& &  & \qty{1080}{\pixel} &  60  & & &  & &  \\
\hline %done

\multirow{2}{*}{\cite{mercat2017energy}} & 
Software energy   & \multirow{2}{*}{HEVC (Kvazaar)} & \qty{240}{\pixel}, \qty{480}{\pixel},  & \multirow{2}{*}{--}& \multirow{2}{*}{--}& 22, 27, & \multirow{2}{*}{PC} & \multirow{2}{*}{Intel CPU} & \multirow{2}{*}{RAPL~\cite{khan2018rapl}} \\
& analysis &  & \qty{720}{\pixel}, \qty{1080}{\pixel} & & & 32, 37 &  & & \\
\hline

\multirow{3}{*}{\cite{uitto_energy_2016}} & Software energy & AVC (x264) & \multirow{3}{*}{\qty{1080}{\pixel}}& \multirow{3}{*}{ \num{25}} & \multirow{3}{*}{\texttt{Ultrafast}} &\multirow{3}{*}{--} & \multirow{3}{*}{PC} &\multirow{3}{*}{Intel CPU} &  Eaton Managed  \\
& analysis & HEVC (x265, Kvazaar) & & & & &  & & ePDU~\cite{eaton} \\
&  & VP9 (Vpxenc) & & & & & & & \\
\hline %done

\multirow{4}{*}{\cite{katsenou_energy-rate-quality_2022}} & & AV1 (SVT-AV1) & \qty{416}{}$\times$\qty{240}{}, & \multirow{2}{*}{30, 50,} &  & & \multirow{4}{*}{PC} & \multirow{4}{*}{Intel CPU} & \multirow{4}{*}{RAPL~\cite{khan2018rapl}}  \\
 & Software energy  & VVC (VVenC) & \qty{832}{}$\times$\qty{480}{}, & \multirow{2}{*}{60}& \multirow{2}{*}{--} &\multirow{2}{*}{--} &  & &  \\&  analysis & VP9 (VP9)&\qty{1080}{\pixel} & & & & & &  \\ &  & HEVC (x265)& &   & & & & &  \\
\hline

\multirow{2}{*}{\cite{sharrab_aggregate_2013}} & {Linear software} & AVC (x264) & \qty{160}{}$\times$\qty{120}{}, & \multirow{2}{*}{\numrange[range-phrase=--]{1}{30}} &  \multirow{2}{*}{--}& \multirow{2}{*}{\numrange[range-phrase=--]{1}{30}}& \multirow{2}{*}{Laptop}& \multirow{2}{*}{Intel CPU} & \multirow{2}{*}{Pro ES AC~\cite{ramseyer2013watts}} \\
&  power model & MPEG-4 Part 2 (FFmpeg) &  \qty{600}{\pixel} &  & & & & & \\
\hline %done

\multirow{2}{*}{\cite{alaoui2014energy}} & Exponential software    & \multirow{2}{*}{AVC (JM)} & \qty{352}{}$\times$\qty{288}{}, &  \multirow{2}{*}{\numrange[range-phrase=--]{1}{30}} &  \multirow{2}{*}{--} & \multirow{2}{*}{\numrange[range-phrase=--]{0}{51}}& \multirow{2}{*}{PC} & \multirow{2}{*}{Intel CPU} & \multirow{2}{*}{--}\\
 &  energy model & &\qty{176}{}$\times$\qty{144}{}  &  & &   &  &   & \\
\hline %done
\multirow{2}{*}{\cite{seeliger2022green}} & Hardware
energy  &  AVC (FFmpeg) & \qty{720}{\pixel},  \qty{1080}{\pixel},& \multirow{2}{*}{25} & \multirow{2}{*}{--}& \multirow{2}{*}{--}& \multirow{2}{*}{PC} & \multirow{2}{*}{--} & \multirow{2}{*}{Power meter}  \\ 
& analysis  and AI model &HEVC (FFmpeg)& \qty{2160}{\pixel}& & & &  & & \\
\hline

\multirow{2}{*}{\cite{amirpour_optimizing_2023}} &  ML-based software
 &  \multirow{2}{*}{HEVC (x265)} & \multirow{2}{*}{\qty{2160}{\pixel}}  &\multirow{2}{*}{30} & \texttt{Placebo} --&\multirow{2}{*}{--} & Data & \multirow{2}{*}{Intel CPU} & \multirow{2}{*}{CodeCarbon~\cite{codecarbon}}\\
& energy model & &  & & \texttt{ultrafast}& \multirow{2}{*}{--}& center &  & \\
\hline
\end{tabular}
}
\end{table*}

\paragraph*{Codec} 
As the processing complexity of video encoding increases, the energy consumption also rises accordingly~\cite{silveira_performance_2017}. The computational requirements for performing intricate algorithms and calculations, especially in high-resolution videos, result in higher power use primarily attributed to the intensive computational tasks. 

\begin{description}[font=\normalfont\itshape]
\item[AVC versus old encoders.]
Sharrab~\etal~\cite{sharrab_aggregate_2013} showed that AVC consumes over four times more power than earlier standards like MJPEG and MPEG-4 Part 2 due to different compression techniques. Specifically, AVC allows multiple reference frames, which increases the compression ratio but also consumes more power. On the other hand, MJPEG only uses intra-frame compression and does not use reference frames, while MPEG-4 allows for two reference frames.

\item[AVC versus HEVC.]
Monteiro~\etal~\cite{monteiro_rate-distortion_2015} compared the compression efficiency and energy consumption of AVC and HEVC reference implementations (\ie JM and HM). The experimental results revealed a superior compression efficiency of HEVC by \qty{25.1}{\percent} with an increased energy consumption of \qty{17.4}{\percent} compared to AVC.  Additionally, the difference in energy consumption between HM and JM increases as QP decreases. However, the HM encoder demonstrated favorable results at QP 37, consuming less energy than JM in some classes. Thus, HEVC is a better energy and compression efficiency choice for high QP values like 37. 

\item[AVC, HEVC, VP9, and AV1.]
The evaluation study in~\cite{uitto_energy_2016} uses open-source software encoders~(\ie x264, x265, VP9) instead of their reference software and showed that x264 had the lowest energy consumption but the worst compression efficiency. In contrast, x265 had the best compression efficiency but higher energy consumption. The VP9 encoder provided the best balance between compression efficiency and energy consumption. A study on the energy consumption of four state-of-the-art video encoders~\cite{katsenou_energy-rate-quality_2022} (\ie HEVC, VVC, VP9, AV1) using their open-source software implementations (\ie x265, VVenC, VP9, SVT-AV1) showed that SVT-AV1 offers the best trade-off between quality, bitrate, and energy. Meanwhile, x265 performed the best for low-energy solutions, although with slightly lower quality.
\end{description}

\paragraph*{Encoding configuration}
This paragraph reviews the impact of encoding parameters on energy.

\begin{description}[font=\normalfont\itshape]
\item[Quantization Parameter (QP) and framerate.]
Reducing framerate and increasing QP lead to reduced energy consumption~\cite{alaoui2014energy} in AVC. QP has an exponential impact on power consumption~\cite{sharrab_aggregate_2013}.

\item[Inter- and intra-prediction.] The work in~\cite{sharrab_aggregate_2013} emphasized that inter- and intra-predictions consume significantly more energy than tasks such as transform, quantization, entropy coding, and decoding in AVC. Additionally, the complexities of inter-prediction, intra-prediction, rate-distortion optimization mode selection, and sub-pixel search are primarily influenced by temporal and spatial resolutions, often called pixel rate.

\item[Preset.]
Faster presets in video encoding reduce energy consumption due to their streamlined computational processes. These presets use less intensive algorithms, simpler motion estimation methods, and less precise rate control, sacrificing some compression efficiency for faster encoding. The work in~\cite{silveira_performance_2017} evaluated the impact of x265 presets using ten x265 encoding presets ranging from \texttt{ultrafast} to \textit{placebo}, trading off the encoding speeds for compression. The \texttt{ultrafast} preset had a \qty{45}{\percent} higher bitrate and 45-fold higher energy consumption than the \texttt{placebo}, indicating a \qty{145}{\percent} increase in energy for every \qty{1}{\percent} reduction in bitrate. 

\item[Resolution.]
Mercat~\etal~\cite{mercat2017energy} provided an analysis of energy reduction techniques achieved by HEVC for real-time encoders at various resolutions and granularity levels. They found that energy savings are directly proportional to video resolution at a coarse granularity, while transform skipping proved ineffective at a middle granularity. At a lower granularity, the coding tree unit level showed the potential for substantial energy reduction, up to \qty{78.1}{\percent}, in contrast to the intra-prediction level, which offered a maximum of \qty{30}{\percent} energy reduction.

\item[Other encoding parameters.]
Encoding parameters such as the number of reference frames, search range, sub-pixel search range, and ME algorithm can lead to power consumption variances of up to \qty{10}{\percent}~\cite{sharrab_aggregate_2013}. Menteiro~\etal~\cite{monteiro_rate-distortion_2015} evaluated the impact of the most computation-intensive process ME on the HEVC performance and observed that, while the ME range has a minor impact on compression efficiency, the energy consumption of HEVC increases more rapidly with increasing ME range for the same video quality. 
\end{description}

\paragraph*{CPU and memory energy use}
An open-source implementation of HEVC in~\cite{silveira_performance_2017} showed that the CPU accounted for \qty{95}{\percent} of the total energy consumption of the x265 systems. In contrast, DRAM represented \qty{3}{\percent}, and the cache memories accounted for \qty{2}{\percent} for the tested sequences.  

\paragraph{Energy-aware video encoding} 
\label{sec:encoding-aware}
Several works proposed models and techniques to estimate or reduce the encoding energy consumption, as presented in Table~\ref{tbl:encoding}.  

Sharrab~\etal~\cite{sharrab_aggregate_2013}  estimated the power consumption in AVC for live video streaming by representing the computational complexity as a linear function of the pixel rate derived from the spatial and temporal resolutions of the raw video. The analysis concluded that parameter tuning often involves a delicate balance between energy consumption and bitrate, and consequently, it developed models to account for the bitrate in this trade-off. 
Alaoui~\etal~\cite{alaoui2014energy} proposed a model to predict energy consumption for AVC for constrained networks, such as wireless video sensor networks, considering the QP and FR parameters when employing intra-image coding only.

Seeliger~\etal~\cite{seeliger2022green} introduced an artificial intelligence (AI) content-aware encoding method named DeepEncodet that predicts the quality of each scene and recommends optimized encoding settings, including varying bitrates and resolutions.
%Measurements demonstrate the effectiveness of DeepEncode in optimizing resolutions and bitrates, leading to a substantial reduction in power usage.
On average, DeepEncode reduces power consumption by \qty{3}{\watt} representing \qty{9}{\percent}) for HD videos and by \qty{69}{\watt} representing \qty{30}{\percent} for UHD videos compared to the standard FFmpeg software encoder.

Amirpour~\etal~\cite{amirpour_optimizing_2023} explored the energy consumption versus quality trade-off in HEVC video encoding, explicitly using the x265 encoder, and revealed that transitioning to slower presets at higher bitrates notably increased energy consumption with a marginal improvement in video quality. Leveraging this insight, they studied different machine learning (ML) approaches (\ie random forest, support vector machines, and K-nearest neighbors), recommending specific presets for each bitrate-resolution combination, effectively balancing energy efficiency and video quality and offering a practical solution for optimizing video encoding configurations.

\paragraph{Video encoding codecs distribution}  
\begin{wrapfigure}{R}{0pt}
    \centering
    \includegraphics[width=.4\columnwidth]{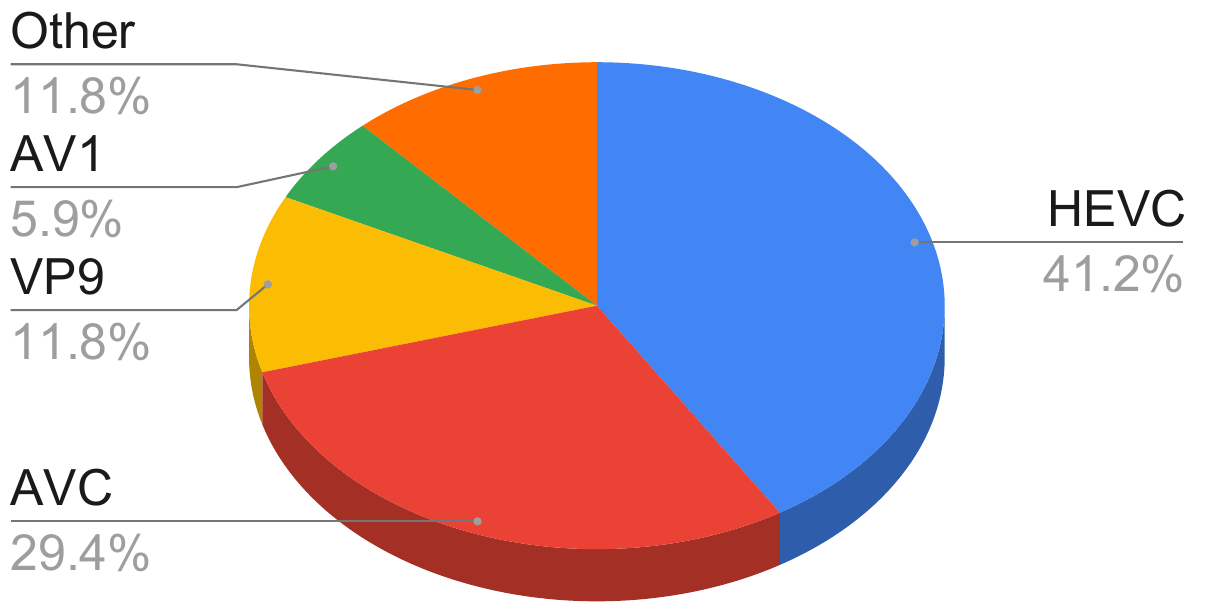} 
    \caption{\centering Distribution of codecs utilized in the related works %classified 
    for video encoding in Table~\ref{tbl:encoding}.}
    \label{fig:Enc_codec} 
\end{wrapfigure}

Table~\ref{tbl:encoding} shows study efforts on the energy consumption for different video encodings (see Sections~\ref{sec:encoding-analysis} and~\ref{sec:encoding-aware}), with high attention dedicated to AVC~\cite{AVC} and HEVC~\cite{sullivan2012overview}. Additional investigations explored energy consumption for alternative codecs such as VP9~\cite{vp9} and AV1~\cite{han_technical_2021}, while the landscape of VVC~\cite{bross_overview_2021} remains an area with limited research (see Figure~\ref{fig:Enc_codec}).

\subsubsection{Resource requirements in content provisioning}
\label{sec-cc}
% \subsection{Energy efficiency in computing continuum}

% Tseng, T. Yang, K. Yang and P Chen, An Energy Efficient VM Management Scheme with Power-Law Characteristic in Video Streaming Data Centers”, IEEE Transactions on Parallel and Distributed Systems, vol. 29, no.2, pp. 297-311, 2018.

Recently, the \emph{computing continuum}~\cite{balouek2019towards} federating the cloud infrastructures with emerging edge devices for video encoding services~\cite{Afzal2022} gained considerable attention. In this section, we explore the studies on video encoding in the computing continuum, summarizing the research works based on the optimization methods and core algorithms used, such as linear programming, heuristic, parallelization, system dynamics, and game theory (see Table~\ref{tbl:related:compcont}).

\begin{table*}[!t]
\caption{Related work classification of video encoding in the computing continuum.}
\label{tbl:related:compcont}

\resizebox{\textwidth}{!}{
\begin{tabular}{|l|l|l|l|l|l|l|l|l|l|l|l|}
\hline
\multirow{2}{*}{\emph{Work}} & \multirow{2}{*}{\emph{Goal}} & \multirow{2}{*}{\emph{Method}} &  \multirow{2}{*}{\emph{Infrastructure}}&{\emph{Energy}}& \multicolumn{2}{l|}{\emph{Encoding parameters}} & \multirow{2}{*}{\emph{Experiment type}} & \multicolumn{2}{l|}{\emph{Machine}} & \multicolumn{2}{l|}{ \emph{Energy}}\\
\cline{6-7} \cline{9-12}

& & & &{\emph{type}}&{\emph{CD}} & {\emph{RS [pixels]}}&  & {\emph{Type}} & {\emph{Processor}} & {\emph{Tool}}& {\emph{Dataset}} \\
\hline\hline

\multirow{2}{*}{\cite{lim2017efficient}} & Max. utilization   & \multirow{2}{*}{Heuristic} & \multirow{2}{*}{Cloud}  & \multirow{2}{*}{--}&\multirow{2}{*}{AVC}&   \multirow{2}{*}{\SI{1080}{\pixel}} &Simulation & & \multirow{2}{*}{--} &  \multirow{2}{*}{--}&  \multirow{2}{*}{--} \\
& Min. storage & &  & & & & Real& PM, VM & &  &  \\
\hline

\multirow{2}{*}{\cite{lee2019video}}& Max. video quality & \multirow{2}{*}{Heuristic} &\multirow{2}{*}{Cloud}  &  \multirow{2}{*}{--}& \multirow{2}{*}{--} & \SI{240}{\pixel}, \SI{360}{\pixel}, &Simulation & \multirow{2}{*}{PC} & \multirow{2}{*}{Intel CPU}   &\multirow{2}{*}{--}  &\multirow{2}{*}{Power~\cite{song2015scheduling}}\\
&  Min. energy & &   & & & \SI{480}{\pixel}, \SI{720}{\pixel} &   Real & &   & &  \\
\hline

% Batmunkh\cite{batmunkh2022carbon} & Cloud & Min. energy &  Heuristic &  &2160p& & & &Real & & & & \\
% & & &   &  & & & & & & & & & \\
% \hline

\multirow{2}{*}{\cite{n2022greenfog}} & \multirow{2}{*}{Min. energy} &  \multirow{2}{*}{LP}& \multirow{2}{*}{Edge} &\multirow{2}{*}{Renewable}& \multirow{2}{*}{--} &\multirow{2}{*}{\SI{720}{\pixel}} &\multirow{2}{*}{Real} & VM &--&Eaton Managed &  Renewable \\
& & &     & & & & &SBC & ARM CPU & ePDU~\cite{eaton} &energy~\cite{toosi2017renewable}\\
\hline

\multirow{3}{*}{\cite{mehrabi2019energy}} & Min. energy &  \multirow{3}{*}{INLP} &\multirow{3}{*}{Edge} &   \multirow{3}{*}{Renewable}&\multirow{3}{*}{--}& \multirow{3}{*}{--}& Simulation & \multirow{3}{*}{--} &\multirow{3}{*}{--} &\multirow{3}{*}{--} & Solar \\
&Max. QoE  &  &   & & & &(SimuLTE~\cite{simulte}) & &  & &energy~\cite{nguyen2018distributed}\\
& Min. network traffic &  &  &  && & & & & & \\
\hline

\multirow{2}{*}{\cite{li2020energy}}&\multirow{2}{*}{Min. energy}&\multirow{2}{*}{MILP} &\multirow{2}{*}{Edge} &\multirow{2}{*}{--} & \multirow{2}{*}{--} & \SI{480}{\pixel}, \SI{720}{\pixel}, & \multirow{2}{*}{Simulation} & \multirow{2}{*}{--}& \multirow{2}{*}{--} &\multirow{2}{*}{--} & \multirow{2}{*}{--}\\
&& & & &  & \SI{1080}{\pixel} &  & &  & & \\
\hline

\multirow{2}{*}{\cite{li2017leveraging}} &Min. energy, &\multirow{2}{*}{Heuristic} &\multirow{2}{*}{Edge} &\multirow{2}{*}{Renewable}& \multirow{2}{*}{AVC} &\SI{360}{\pixel}, \SI{480}{\pixel}, & Real (Grid’5000~\cite{balouek2013adding}) & \multirow{2}{*}{VM} & \multirow{2}{*}{Intel CPU}  &\multirow{2}{*}{--} &Solar  \\
& Max. QoE  & &   & & & \SI{720}{\pixel}&  Simulation~\cite{li2015opportunistic} &  &  && power~\cite{photovolta2}\\
\hline

\multirow{2}{*}{\cite{pradeepkumar2021resource}} &\multirow{2}{*}{Min. network traffic} %frame trans.
&\multirow{2}{*}{Heuristic} &\multirow{2}{*}{Edge} &\multirow{2}{*}{--} & \multirow{2}{*}{AVC} & \SI{480}{\pixel}, \SI{540}{\pixel},&  \multirow{2}{*}{Real}&\multirow{2}{*}{PC}&\multirow{2}{*}{Intel CPU} &\multirow{2}{*}{Power meter} & \multirow{2}{*}{--}\\
& &  &  & & & \SI{720}{\pixel} & & & & & \\
\hline

\multirow{2}{*}{\cite{beck2015me}} &\multirow{2}{*}{Min. energy}& \multirow{2}{*}{Heuristic} &\multirow{2}{*}{Edge} &\multirow{2}{*}{Renewable}&\multirow{2}{*}{AVC} &\SI{144}{\pixel}%176$\times$144, %CIF and QCIF https://qualinet.github.io/databases/video/video_trace_library_asu/
&\multirow{2}{*}{Real}&\multirow{2}{*}{Smartphone}&\multirow{2}{*}{--}& %Power meter??
\multirow{2}{*}{--}& Smartphone's \\
& & &  & & & \SI{288}{\pixel}%352$\times$288
&  & &   & & energy~\cite{lauridsen2014empirical}\\

\hline

\multirow{2}{*}{\cite{liu2016greening}} & \multirow{2}{*}{Min. energy} & \multirow{2}{*}{Heuristic} &\multirow{2}{*}{Edge} &\multirow{2}{*}{--}& \multirow{2}{*}{AVC}&\SI{480}{\pixel}, \SI{720}{\pixel},&\multirow{2}{*}{Real}& \multirow{2}{*}{SBC} &ARM CPU&\multirow{2}{*}{--}&\multirow{2}{*}{--}\\
& &  & %(Transcoder development)
&&  &1080 &  & &Broadcom GPU  & & \\
\hline

\multirow{2}{*}{\cite{bernabe2018parallel}} &\multirow{2}{*}{Min. energy}&\multirow{2}{*}{Parallelization} &\multirow{2}{*}{Edge} & \multirow{2}{*}{--}
&MPEG-4 & \multirow{2}{*}{\SI{512}{}$\times$\SI{512}{}}&
\multirow{2}{*}{Real} & \multirow{2}{*}{SBC} & \multirow{2}{*}{ARM CPU} & \multirow{2}{*}{--} & \multirow{2}{*}{--}\\
&  & &   & & part 1 and 2 & & &  & & & \\
\hline

\multirow{2}{*}{\cite{rigazzi2019edge}} & Min. power, &\multirow{2}{*}{Heuristic} &Cloud,&\multirow{2}{*}{--}&\multirow{2}{*}{HEVC}& \multirow{2}{*}{\SI{2160}{\pixel}} &
Real & Cloud: PC &  Nvidia GPU&\multirow{2}{*}{GPU-z~\cite{techpowerup}}&\multirow{2}{*}{--}\\
&Min. bandwidth &  & Edge & &  & & (Fog05~\cite{eclipsefog05})& Edge: Laptop &  Intel CPU& & \\
\hline

\multirow{2}{*}{\cite{mekonnen2017energy}} & \multirow{2}{*}{Energy analysis} & \multirow{2}{*}{Heuristic} &Cloud &\multirow{2}{*}{--}
&\multirow{2}{*}{AVC}& \multirow{2}{*}{\SI{480}{\pixel}} &Simulation & Cloud: SBC & \multirow{2}{*}{ARM CPU} & Monsoon power &\multirow{2}{*}{--}\\ 
&  &  & Edge& & & &  Real & Edge: SBC  &  
% ARM Cortex-A7 CPU, quad-core ARM cortex A53 CPU
& meter~\cite{monsoonpowermonitor}& \\
\hline

\multirow{2}{*}{\cite{ramprasad2021sustainable}} &\multirow{2}{*}{\ch{CO2} analysis}&\multirow{2}{*}{SD}& Cloud,  &\multirow{2}{*}{Renewable}&\multirow{2}{*}{AVC}&\multirow{2}{*}{\SI{1080}{\pixel}, \SI{2160}{\pixel}}&
Simulation&\multirow{2}{*}{PM}&\multirow{2}{*}{Intel CPU}&\multirow{2}{*}{EPA~\cite{usa-epa-ghg}}&\multirow{2}{*}{\ch{CO2}~\cite{usa-epa-ghg,dell-PowerEdge}}\\
& & & Edge & &  & &(Vensim~\cite{vensim})  &  & & &\\
\hline

\multirow{2}{*}{\cite{gmsys2023afzal}}& Min. energy, & \multirow{2}{*}{Game theory}  & Cloud,  & \multirow{2}{*}{Renewable} & \multirow{2}{*}{HEVC} & \multirow{2}{*}{\SI{2160}{\pixel}} & \multirow{2}{*}{Real}   & Cloud: AWS VM & Intel CPU& \multirow{2}{*}{CodeCarbon \cite{codecarbon}}&Electricity\\
 & Min. compute cost & & Edge&  &  & & & Edge: PC  & Intel CPU& & Maps~\cite{tranberg2019real,elecmap}\\
\hline
\end{tabular}}
\end{table*}

\paragraph*{Heuristic} methods offer practical suboptimal solutions in a polynomial time~\cite{afzal2022otec}.
Lim~\etal~\cite{lim2017efficient} proposed a resource consolidation algorithm for allocating physical machines (PMs) to virtual machines (VMs) and VMs to video tasks. Lee~\etal~\cite{lee2019video} presented a heuristic algorithm for selecting transcoding parameters that reduces power consumption. Li~\etal~\cite{li2017leveraging} proposed an offloading method for video encoding applications to the edge or the cloud, depending on renewable energy availability. Beck~\etal~\cite{beck2015me} introduced an edge offloading method to reduce the power consumption of mobile devices. Liu~\etal~\cite{liu2016greening} introduced an energy-efficient video transcoding cluster with low-cost single-board computers. Rigazzi~\etal~\cite{rigazzi2019edge} proposed a microservices-based design for distributing \SI{360}{\degree} video streaming services, while Mekonnen~\etal~\cite{mekonnen2017energy} investigated the impact of virtualization on battery life.

\paragraph*{Linear programming (LP)} is a mathematical optimization finding the near-optimal solution considering a set of constraints~\cite{afzal2022otec}. However, it can increase the search space.
Toosi~\etal~\cite{n2022greenfog} proposed an LP method for optimizing the framerate and resolution of video streaming based on the available renewable power and minimizing non-renewable energy consumption on low-cost single-board computers (SBC).  % Edge
Mehrabi~\etal~\cite{mehrabi2019energy} designed an integer non-LP optimization method to tune the network traffic and quality of experience in dynamic adaptive video streaming over HTTP (DASH) scenarios for mobile device users supporting collaborative and non-collaborative edge caching and processing by emulating mobile clients' solar power sources and radio access channels. However, it does not evaluate the performance of different video codecs through simulations.  
Li~\etal~\cite{li2020energy} proposed a two-stage stochastic mixed-integer LP (MILP) for video caching and delivery that minimizes energy consumption to schedule video transcoding on the edge infrastructure. % edge

\paragraph*{Parallelization} divides a computational task into smaller subtasks executed in parallel, which can substantially enhance the speed of task execution, especially for large and complex problems~\cite{Karale2022}. Bernabe~\etal~\cite{bernabe2018parallel} proposed a parallelization strategy for the 3D fast wavelet transform using a cluster of single-board computers, such as edge devices, beneficial for JPEG-2000 image standard compression. This method utilizes the MPEG-4 Part 1 and 2 standards for video compression. The parallelization strategy implemented using the message passing interface (MPI)~\cite{gropp1999using} and POSIX threads~\cite{butenhof1997programming} revealed significant speed improvements on a single Raspberry Pi with up to four times lower energy consumption than the higher-performance 4-core Intel Xeon processors. However, spreading all MPI processes across multiple boards drops performance due to the limited bandwidth of the onboard LAN port, which is insufficient for the high-volume communication demands.

\paragraph*{System dynamics (SD)} is a method to model, simulate, analyze, and design complex systems that change over time~\cite{sterman2002system,palm2010system}. 
Ramprasad~\etal~\cite{ramprasad2021sustainable} used SD modeling to estimate the \ch{CO2} footprint caused by the mobile wireless infrastructure and video application requests on cloud servers spread across the wide area network in the short and long term.  The analysis highlighted that the base station radio and the wide area network contribute significantly to \ch{CO2} emissions. To further reduce them, Ramprasad~\etal investigated the impact of deploying applications on the cloud and edge, considering each service provider's energy efficiency and renewability rate. They found that placing edge data centers near base stations and using new 5G mobile network features potentially reduces \ch{CO2} emissions by up to \SI{50}{\percent}, compared to uploading data from smart cameras to the cloud and base station not configured to ``power off'' mode.

\paragraph*{Game theory} studies strategic decision-making involving two or more interdependent stakeholders~\cite{owen2013game}. Afzal~\etal\cite{gmsys2023afzal} proposed a matching game scheduler on edge and cloud computing instances that models a video encoding application consisting of codec, bitrate, and resolution set. They achieved 
%reduction in video encoding cost expenses by \qtyrange{17}{78}{\percent} in the cost-prioritized matching-based method, while a decrease of 
a \SIrange{38}{45}{\percent} decrease in energy use and \SI{80}{\percent} in \si{\gram}\ch{CO2} emissions of the energy-prioritized method compared to a cost-prioritized one. %Hybrid

\begin{table}[!t]
\caption{Related work classification on encoded video content storage (BR: bitrate).}
\label{tbl:related:cdn}
\resizebox{\columnwidth}{!}{
\begin{tabular}{|l|l|l|l|l|l|l|}
\hline
\multirow{2}{*}{\emph{Work}} & \multirow{2}{*}{\emph{Goal}} & \multirow{2}{*}{\emph{ Method}} &  \multirow{2}{*}{\emph{Energy type}} & \multicolumn{2}{l|}{\emph{Encoding parameters}} & \multirow{2}{*}{\emph{Experiment type}}\\
\cline{5-6}
&& && \emph{CD} & \emph{BR [Mb/s]}  & \\
\hline\hline

\multirow{2}{*}{\cite{ul2012evaluating}}&Max. resource &\multirow{2}{*}{Heuristic} &\multirow{2}{*}{--}& \multirow{2}{*}{--}& \multirow{2}{*}{--}& Simulation \\ 
  &utilization &  && & & CDNsim~\cite{stamos2010cdnsim},  OMNeT++~\cite{omnetpp} \\
  \hline
 
\multirow{2}{*}{\cite{Seeliger23End-to-End}}&\multirow{2}{*}{Min. energy}&\multirow{2}{*}{Heuristic}&\multirow{2}{*}{Renewable}& \multirow{2}{*}{--}& \multirow{2}{*}{--}&\multirow{2}{*}{--} \\ 
 &  &  & & && \\
 \hline
 
\multirow{2}{*}{\cite{osman2014energy}}&Min. energy,&\multirow{2}{*}{MILP}&\multirow{2}{*}{--} &\multirow{2}{*}{MPEG-4, AVC} & \numrange[range-phrase=--]{1.5}{3}, &  Simulation \\ 
 &   Min. traffic&  &&  & \numrange[range-phrase=--]{8}{12} & CDNsim~\cite{stamos2010cdnsim}, OMNeT++~\cite{omnetpp} \\
 \hline

\multirow{2}{*}{\cite{goudarzi2019joint}}&Min. energy, & LP, DE, &\multirow{2}{*}{Renewable}&\multirow{2}{*}{MPEG-4,  AVC}&\multirow{2}{*}{\num{0.512}} & Simulation \\
 &  Max. QoE& game theory & &  & & CDNsim~\cite{stamos2010cdnsim}, OMNeT++~\cite{omnetpp} \\

 \hline
\end{tabular}}
\end{table}

\subsection{Storage}
%Content Delivery Networks (CDNs) play a pivotal role in storing encoded video content. 
CDNs of strategically positioned distributed cache servers based on their geographical location closer to the users improve the performance of Internet video services by reducing delivery delays and bandwidth consumption with energy consumption trade-offs. A CDN stores the entire dataset and several surrogate machines, each caching a subset of the entire dataset. The optimal caching strategy depends on the video content's popularity, size, quality, network topology, and traffic patterns. Therefore, it is essential to study and design energy-efficient caching algorithms that balance the benefits and costs of caching. Table~\ref{tbl:related:cdn} summarizes recent studies on storing encoded video on CDN, categorized into heuristic, LP, differential evolution (DE), and game theory.

\paragraph*{Heuristic}
Islam~\etal~\cite{ul2012evaluating} developed an energy consumption model for surrogate servers using CDNs. While decreasing the number of surrogate servers can reduce CDN energy consumption, its increased transport cost leads to higher overall CDN energy costs. Moreover, a decrease in the number of surrogates may result in increased mean response time, decreased byte-hit ratio, and increased failed requests attributed to the increased load. While increasing the number of requests may result in a lower mean response time, higher byte-hit ratio, and fewer failed requests, it can also increase the energy for handling the traffic. 
Seelinger~\etal~\cite{Seeliger23End-to-End} proposed an energy-aware method based on multi-CDN content steering, recently adopted in both HTTP live streaming (HLS)~\cite{pantos2022hls} and DASH~\cite{DASH-IF-CTS} standards. In multi-CDN content steering, players use a standard fallback method by analyzing the steering manifest and choosing the designated delivery channel when a data center is down or overloaded. Content providers often switch CDNs using streaming analytics, considering QoE measures like average bitrate, warning and error rates, and throughput. Consequently, the ranking of CDNs in the steering manifests may constantly change to direct players toward a new CDN, considering cost and energy preferences and prioritizing green delivery routes powered by renewable energy sources. Integrating this information with QoE measures decides the order of CDNs in the content steering manifests. End-consumers are unaware of this smooth transition between CDNs due to recent HLS and DASH standards updates.

\paragraph*{Linear programming}
Osman~\etal~\cite{osman2014energy} addressed the issue of reducing the high energy consumption of delivering Internet video content by developing a MILP method in a core network that deploys the Internet protocol over wavelength division multiplexing and video cache and minimizes power consumption by optimizing the size of caches on the servers.

\paragraph*{Differential evolution}
Differential evolution (DE)~\cite{price2006differential} is a stochastic algorithm that seeks a function's global optimum, falling under the broader category of evolutionary algorithms, inspired by the principles of natural selection.
Goudarzi~\etal~\cite{goudarzi2019joint} proposed a joint optimization method to maximize end-users QoE and minimize the operational cost of electricity based on LP-optimization and nonlinear DE methods assuming a different number of renewable solar energy sources distributed uniformly among CDNs. While LP offers a low complexity for joint optimization problems, DE achieves the global optimal with a higher energy cost. Both solutions offer a trade-off between operational cost savings and complexity, which varies with the total number of solar panels.

\paragraph*{Game theory} Goudarzi~\etal~\cite{goudarzi2019joint} proposed a game-theoretic method with CDN providers making independent decisions and attempting to optimize their payoff unilaterally satisfied at a Nash equilibrium. With this game-theoretic model, CDN providers can exchange traffic and connections based on the availability of their green energy sources.

\subsection{Research gap}
\label{sec:rg_enc}

% https://ieeexplore.ieee.org/stamp/stamp.jsp?arnumber=8481649&tag=1

\paragraph*{Optimized video encoding parameters}
Optimizing the video encoding parameters presents a significant potential to reduce energy consumption while maintaining high-quality output by strategically adjusting parameters such as framerate, resolution, and bitrate.
% video codec, resolution, framerate, and bitrate, it is possible to lower the bitrates of encoded videos without compromising their quality. This approach also includes considering power-efficient codecs and other relevant parameters to optimize overall energy in video streaming.

\paragraph*{Energy efficient transcoding} Accelerating video content transcoding with less energy consumption at an acceptable quality for the end-users~\cite{de2015fast, amirpour_between_2022} relies on extracting valuable information from the decoding process. For example, HEVC-to-VP9 transcoding achieved significant time reductions and energy savings by using information extracted from the HEVC decoder~\cite{de2015fast} and discarding certain coding modes during VP9 encoding.

\paragraph*{On-demand HLS segment encoding and storage}
HLS typically encodes video segments at a fixed set of bitrate-resolution pairs, forming a bitrate ladder. However, end-users~\cite{quortex2022mission} never request a substantial number of segments, as encoding and storing segments without any corresponding benefit requires significant processing. A potential solution is to encode and package segments on-demand based on user requests rather than encoding and transmitting the entire ladder to the edge or CDN to reduce the energy consumption~\cite{gmsys2023studying}. 
In addition, the energy consumption of non-requested segments on CDN is another critical factor, addressed by new techniques, such as content caching upon popular requests and the lightweight transcoding approach~\cite{erfanian2021lwte}

\paragraph*{Optimized HLS video bitrate ladder}
Generating a bitrate ladder in HLS is highly power-intensive, resulting in high operational costs. Eliminating perceptually redundant representations comparing the video multimethod assessment fusion (VMAF) scores of two representations (\ie below a threshold~\cite{yuan2019visual}) and removing the higher bitrate representation if it is perceptually lossless significantly reduces energy consumption and storage usage and, consequently, the carbon footprint and operational costs~\cite{menon2023content}. Alternatively, different types of content have varying bitrate requirements~\cite{koziri2018efficient}, such as cartoons compressable with acceptable quality at significantly lower bitrates. Considering user-perceived quality metrics can decrease the bitrates of encoded videos while maintaining their quality and reducing the energy consumption of video streaming. 
%Video compression with lower bitrates can decrease the energy consumption of video streaming.

\paragraph*{Optimized ABR multi-codec bitrate ladder} 
Over time, streaming systems have adapted to incorporate a variety of codecs. Older devices primarily rely on AVC, while newer ones may use HEVC or AVC streams. While advanced devices support multiple codecs~\cite{yuriy_multicodec_ref}, managing them involves creating ABR bitrate ladders for each codec separately, encoding and storing all segments into multiple representations with different bitrate and resolution combinations, leading to a significant computational workload and energy use~\cite{koziri2018efficient}. A multi-codec bitrate ladder that removes unnecessary high-bitrate representations of new-generation codecs in case of negligible quality difference can substantially reduce energy consumption for the streaming system~\cite{menon2023energy}.

\paragraph*{Cloud resources profiling}
An energy profiler providing fine-grained power consumption and GHG emission information, including energy labels (\eg A+, A, B, C) for a combination of cloud instance resources and video encoding algorithms compliant to the EU legislation~\cite{eu_energy_labels} requires further research.

\paragraph*{Energy-based pricing for cloud services} A potential improvement in cloud pricing mechanisms could involve energy-based pricing options, such as charging per joule instead of by hour per core.

\paragraph*{Optimized virtualized cloud technology} Monitoring energy consumption is more challenging for VMs running on the public cloud infrastructure than on dedicated physical machines since monitoring their energy use is not controlled by the service customers. 

\paragraph*{Energy efficient cloud services} Environmentally friendly and energy-efficient cloud services contributes to reducing \ch{CO2} emissions. Recently, IT service providers have been focusing on developing energy-efficient hardware, such as tensor processing units, high-performance servers, and ML algorithms that automatically optimize cooling systems. Alongside hardware advancements~\cite{holzle2020datacenters}, it is also crucial to enhance energy efficiency through software development~\cite{leiserson2020there}.

\paragraph*{Low-carbon cloud regions} IT service providers offer cloud computing platforms in multiple regions delivered through a global network of data centers. Various power plants (\eg fuel, natural gas, coal, wind, sun, water) supply them with electricity, generating different amounts of greenhouse gases. Therefore, it is essential to consider the carbon emitted by power plants in the selected region for cloud computing and their source of energy production.

\paragraph*{Reduced power overhead} The power consumption of idle instances, or ``power overhead''~\cite{davy2021estimating} significantly contributes to the unnecessary carbon footprint due to over-provisioning for potential worst-case scenarios, such as high user demands. Data centers and network security measures contain excessive redundancies, such as redundant power supplies, networks, and equipment constantly running in standby mode, increasing electricity consumption. Data centers prioritize uptime and reliability, and thus, only use \qtyrange{6}{12}{\percent} of their electricity for actual computations, the rest remaining idle for sudden activities~\cite{horner2016known}.

\paragraph*{Green edge services} Recently, the edge services close to end-users in the computing continuum can take advantage of less content transmission, leading to lower \ch{CO2} emissions. Along with renewable power plants at cloud premises, edge providers are using green energy sources, especially in geographical regions that rely on renewable energy production from its related materials such as solar, wind, hydro, geothermal, and biomass~\cite{gmsys2023afzal}.

\begin{table*}[!t]
\caption{Related work classification of content retrieval on end-user devices.}
\label{tbl:related:recieve}
\resizebox{\textwidth}{!}{
\begin{tabular}{|@{ }l@{ }|@{ }l@{ }|@{ }l@{ }|@{ }l@{ }|@{ }l@{ }|@{ }l@{ }|@{ }l@{ }|@{ }l@{ }|@{ }l@{ }|@{ }l@{ }|@{ }l@{ }|}
\hline
\multirow{2}{*}{\emph{Work}}& \multirow{2}{*}{\emph{Goal}} & \multirow{2}{*}{\emph{Method}} & \multicolumn{2}{l|}{\emph{Encoding parameters}} & {\emph{Experiment}}& \multicolumn{2}{l|}{\emph{Machine}} &\emph{Network}& \multirow{2}{*}{\emph{Player}}&{\emph{Energy}}\\
\cline{4-5} \cline{7-8}

 & &&{\emph{CD}} & {\emph{RS}}& \emph{type} & {\emph{Type}} & {\emph{Brand}}& {\emph{protocol}} & &\emph{tool}\\
\hline\hline

\multirow{2}{*}{\cite{zhang2016profiling}}& LTE  energy &\multirow{2}{*}{Profiling}&\multirow{2}{*}{--} & \qty{360}{\pixel}, \qty{720}{\pixel},
 &\multirow{2}{*}{Real}& \multirow{2}{*}{Smartphone}&Samsung Galaxy S5, & \multirow{2}{*}{--} &\multirow{2}{*}{DASH-IF~\cite{dashif-reference-player}}&Monsoon power\\
&  model&& & 
\SI{1080}{\pixel} &  & &Note 3  & & & monitor \cite{monsoonpowermonitor}\\
\hline

\multirow{3}{*}{\cite{huang2012close}} &  \multirow{2}{*}{LTE and WiFi}& \multirow{3}{*}{Linear regression}& \multirow{3}{*}{--} & \multirow{3}{*}{--} &\multirow{3}{*}{Real}  & \multirow{3}{*}{Smartphone}& Motorola Atrix, & \multirow{3}{*}{--} & \multirow{3}{*}{--} & Monsoon power \\
&  \multirow{2}{*}{power model}  & & &  &  & &  Samsung Galaxy S, & & & monitor \cite{monsoonpowermonitor}\\
&   & & &  &  & & iPhone& & & \\
\hline

\multirow{2}{*}{\cite{wei2015power}}  & Min. LTE   &\multirow{2}{*}{Heuristic}& \multirow{2}{*}{--} &  \multirow{2}{*}{--}& \multirow{2}{*}{Real} & \multirow{2}{*}{Smartphone}&\multirow{2}{*}{Google
Nexus 4} & \multirow{2}{*}{--} & \multirow{2}{*}{dash.js~\cite{dashjs}} & \multirow{2}{*}{--}\\
 &  power & & &  &  & & &  &  & \\
\hline

\multirow{2}{*}{\cite{zhang2018optimizing}} & Min. LTE &\multirow{2}{*}{Heuristic}& \multirow{2}{*}{--} &  \SI{180}{\pixel}, \SI{360}{\pixel}, & \multirow{2}{*}{Real} &\multirow{2}{*}{Smartphone} & \multirow{2}{*}{Samsung Galaxy
S5}& \multirow{2}{*}{--} & \multirow{2}{*}{DASH-IF~\cite{dashif-reference-player}} & Monsoon power \\
 &   power & & &  \SI{720}{\pixel},  \SI{1080}{\pixel} &  & & & &  & monitor~\cite{monsoonpowermonitor}\\
\hline

\multirow{4}{*}{\cite{yue2020energy}} & \multirow{2}{*}{LTE and WiFi} & \multirow{4}{*}{Linear regression}& \multirow{2}{*}{AVC,}  & Regular video:   & \multirow{4}{*}{Real} & \multirow{4}{*}{Smartphone}& \multirow{2}{*}{LG V20, }& \multirow{2}{*}{802.11a/b/g,}&\multirow{2}{*}{ExoPlayer~\cite{exoplayer},}  & \multirow{2}{*}{Monsoon power}\\
& \multirow{2}{*}{power  model }&  & \multirow{2}{*}{HEVC}  &\SIrange{144}{1080}{\pixel},   &  & &   \multirow{2}{*}{Moto G5}&  \multirow{2}{*}{802.11n}& \multirow{2}{*}{YouTube} &  \multirow{2}{*}{monitor~\cite{monsoonpowermonitor}}\\
& && &\num{360}\degree ~video:    &  & & & & & \\
& && & \SI{2160}{\pixel} &  & & & & & \\
\hline

\multirow{2}{*}{\cite{sun2014modeling} } &  WiFi energy & \multirow{2}{*}{Linear regression}& \multirow{2}{*}{--}& \multirow{2}{*}{--}&  \multirow{2}{*}{Real} &\multirow{2}{*}{Smartphone}&  \multirow{2}{*}{Nexus S} &\multirow{2}{*}{802.11n}& \multirow{2}{*}{--} & Monsoon power \\
 &  model & & &&   && &&&  monitor~\cite{monsoonpowermonitor} \\
\hline

\multirow{2}{*}{\cite{chowdhury2015http}} & WiFi energy  & \multirow{2}{*}{--} & \multirow{2}{*}{--} & \multirow{2}{*}{--}& \multirow{1}{*}{Real} & \multirow{2}{*}{Smartphone} & \multirow{2}{*}{Galaxy Nexus}& \multirow{2}{*}{--}&\multirow{2}{*}{--}&\multirow{2}{*}{--}\\
&analysis && &  & (Green Miner~\cite{hindle2014greenminer}) & & & & & \\
\hline

\multirow{2}{*}{\cite{DELAOLIVA2013599}}& Ethernet  energy  & \multirow{2}{*}{--} &\multirow{2}{*}{AVC/SVC} & \multirow{2}{*}{--} & \multirow{2}{*}{Real}& \multirow{2}{*}{PC} & \multirow{2}{*}{--}  &{EEE}   & \multirow{2}{*}{--} &\multirow{2}{*}{--}  \\
& analysis &  & &  & & & & (IEEE 802.3az) &&  \\
\hline

\multirow{2}{*}{\cite{vargas2023contribution}} &Ethernet  energy  &\multirow{2}{*}{--}& AVC,  &\multirow{2}{*}{--}&\multirow{2}{*}{Real} & \multirow{2}{*}{PC}& \multirow{2}{*}{--} &  EEE &\multirow{2}{*}{Shaka~\cite{shaka-player-demo}}& 
\multirow{2}{*}{--}\\
& analysis&&  AVC/SVC&  &  & & &(IEEE 802.3az) & & \\
\hline

\end{tabular}}
\end{table*}

\section{Content Consumption}
\label{sec:cc}

Ensuring energy efficiency in end-user devices is critical for the sustainability of video streaming. Three key components significantly affect devices' energy usage~\cite{yue2020energy,turkkan2022greenabr}: \begin{enumerate*} \item content retrieval through the network interface card (NIC), \item video decoding, and \item video display on the end-user screen\end{enumerate*}. Tables~\ref{tbl:related:recieve}, \ref{tbl:related:decod}, and~\ref{tbl:related:disp} provide summaries of recent studies focusing on these three components, closely examined in the following sections.

\subsection{Content retrieval}
The NIC's power consumption depends on several factors, such as the transmission protocol, the network type, the video quality, and the device type~\cite{trestian2012energy, sun2014modeling}. This section explores different network types, including long-term evolution (LTE), wireless fidelity (WiFi), and Ethernet, and compares their energy consumption.

\subsubsection{LTE}
Understanding the impact of video resolution, buffer size, signal strength, and network bandwidth or throughput on energy consumption is vital for making LTE NICs energy-efficient, alongside other efforts, including new energy models.

\paragraph*{ABR segment sizes}
The analysis in~\cite{zhang2016profiling} found an increase in energy consumption with shorter segments due to increased network activity power consumption. For example, \qty{4}{\second} video segments can reduce the average network power by up to \qty{30.71}{\percent} compared to \qty{2}{\second} segments.

\paragraph*{Video resolution}
As the resolution of streaming videos increases, the power consumption of the NIC also increases significantly~\cite{zhang2016profiling}. For instance, increasing the video resolution from \qtyrange{360}{720}{\pixel}, affects the transmission power by approximately \qty{57.4}{\percent} and to \qty{1080}{\pixel} by \qty{87.3}{\percent}.

\paragraph*{Buffer size}
A larger buffer size enables retrieving more video segments using consecutive (or pipelined) segment requests within persistent TCP sessions, leading to lower network power consumption~\cite{zhang2016profiling, zhang2018optimizing}. For example, the player with a buffer size of \SI{60}{\second} can reduce the average network power by \qty{19}{\percent} compared to a \qty{30}{\second} buffer size~\cite{zhang2016profiling}. While a larger buffer size can have benefits, it can also lead to a significant waste of bandwidth and energy~\cite{zhang2018optimizing} in case of inconsistent user behavior prefetching segments in the buffer without watching them.

\paragraph*{Signal strength} 
Signal strength is a crucial factor influencing the energy consumption of LTE NICs~\cite{zhang2016profiling}. Reference Signal Received Power (RSRP) indicates the signal strength a mobile device receives from a cell tower. The Radio Resource Control (RRC) protocol~\cite{3gpp-tr25.813} establishes and terminates connections between the user equipment and the base station. When the signal strength is low, the RRC must transmit more data packets to ensure a reliable communication link, and the NIC needs to decode the weaker signal, both incurring more energy spending. With a strong signal, the RRC operates more efficiently with lower energy consumption.

\paragraph*{Network bandwidth or throughput}
There is a corresponding increase in the power consumption of LTE devices as the data throughput of the downlink or uplink increases~\cite{yue2020energy,huang2012close}. A study in~\cite{huang2012close} found that the throughput-power relationship in different scenarios fits a linear function.

\paragraph*{LTE NIC energy efficiency}
Authors in~\cite{wei2015power, zhang2018optimizing} investigated methods to improve the energy consumption of LTE NIC. Weit~\etal~\cite{wei2015power} implemented a server push-based low-power streaming mechanism in an HTTP DASH video streaming prototype that saved up to \qty{17}{\percent} of battery power. Zhang~\etal~\cite{zhang2018optimizing} introduced an energy-efficient self-adaptive method to exploit the continuous video-watching time to predict users' behavior (\ie with or without skips) and adjust buffer size and the number of fetched segments accordingly.
% Their  using the push technology, where the server actively pushes the following segments without requiring an individual request for each segment. 

\paragraph*{LTE energy model}
%Some efforts focused on developing models to estimate the energy consumption of LTE NICs~\cite{zhang2016profiling, huang2012close}.
Zhang~\etal~\cite{zhang2016profiling} investigated online video streaming using DASH over 4G LTE networks for mobile devices and proposed a power consumption model that considers downlink throughput as the sole factor affecting power consumption without considering RSRP. Modeling the power consumption as a function of downlink throughput and RSRP is challenging and complex due to the difficulty of determining appropriate RSRP ranges. Another study~\cite{huang2012close} developed an empirically derived comprehensive power model of a commercial LTE network considering downlink and uplink throughput. % (for details of the models refer to Appendix~\ref{apendix:lte}). 

\subsubsection{WiFi}
This section explores the energy efficiency of the WiFi NIC during video streaming, covering the impact of the application layer protocol (\ie comparing HTTP/2 and HTTP/1.1), signal strength, and the influence of network bandwidth or throughput on power consumption. Additionally, it discusses existing models for estimating WiFi NIC energy consumption.
% This section analyses the impact of the application
% protocol, signal strength, network bandwidth, and WiFi NIC on the energy consumption. 

% \paragraph*{Impact of different devices}
% According to the study conducted by Sun~\etal~\cite{sun2014modeling}, a minimal variation was observed within individual phones, but significant variation was observed between different phones.

\paragraph*{Application layer protocol}
HTTP/2 is more energy efficient than HTTP/1.1, especially in high-latency scenarios~\cite{chowdhury2015http}. HTTP/1.1 is inefficient due to the overhead of multiple TCP connections, which worsens with longer round-trip time. HTTP/2 uses a single connection with multiplexing to avoid this problem. The benefit of HTTP/2 depends on the round-trip time between the client and the server. Additionally, the encryption required by HTTPS, the standard protocol for user privacy and security, uses more energy than HTTP/1.1.

\paragraph*{Signal strength}
The Received Signal Strength Indicator (RSSI) measures the strength of the signal received by a WiFi device from a WiFi access point. A higher RSSI value indicates a stronger signal~\cite{sun2014modeling}. However, the signal strength alone cannot always capture the wireless channel dynamics. For instance, high energy consumption can occur even with high signal strength due to hidden terminals and interference from the sender side~\cite{ding2013characterizing}.

\paragraph*{Network bandwidth or throughput}
The operational state of a WiFi interface card, active or idle, influences its power consumption. The time spent by the NIC in the active state is a crucial factor impacted by the ratio between the network bandwidth and the average bitrate of a segment, referred to as \emph{slackness}~\cite{yue2020energy}. Higher slackness results in a faster buffer fill rate, leading to shorter duration in the active state and lower average power consumption.
%In the case of ABR, the rate adaptation logic of ABR streaming adjusts to the available network bandwidth, choosing higher or lower-quality segments accordingly. 
%The selection is based on the declared bitrate, which is close to the highest quality. Higher-quality tracks have a bigger difference between peak and average bitrate, resulting in higher slackness and less time spent actively downloading, which leads to lower average power consumption. 
However, increased network bandwidth results in a higher power draw during the active state.

% Several studies have characterized the energy consumption of WiFi networks~\cite{balasubramanian2009energy, ding2013characterizing,li2012energy,pathak2011fine}; however, they do not focus on video streaming.

\paragraph*{WiFi energy model}
%Several efforts developed models to estimate the energy consumption of WiFi NICs~\cite{yue2020energy, huang2012close,sun2014modeling}.
The research in~\cite{yue2020energy} divided downlink throughput into different ranges and developed a linear throughput-based model for each range.  Huang~\etal~\cite{huang2012close} designed a linear power model as a function of the throughput. Sun~\etal~\cite{sun2014modeling} conducted a non-linear relationship between power and throughput, unlike the model in~\cite{huang2012close}, and concluded that a throughput-based model provides high accuracy for a given transport layer protocol and packet size.

\subsubsection{Ethernet}
The IEEE 802.3az Energy Efficient Ethernet (EEE) standard introduces a strategic approach for energy conservation by carefully managing NIC activity modes~\cite{DELAOLIVA2013599}.
%Operating within this standard, the NIC seamlessly transitions into a low-power mode when not actively transmitting data, which is particularly efficient when dealing with bursty Ethernet traffic.

\begin{description}[font=\normalfont\itshape]
\item [Active mode] characterized by the NIC's full operational readiness for data transmission and reception, optimizes the timing of transitions and extends beyond data transfer, effectively reducing the frequency of mode switches and curtailing the additional related energy expenditures.
\item[Sleep mode] (energy-frugal state) strikes an equilibrium between wakeful and dormant periods to capitalize on energy efficiency while considering the modest consumption during wake-up.
\end{description}
%This meticulous synchronization of activity modes underpins the EEE standard's pivotal role in elevating energy efficiency across network devices, thereby underscoring its invaluable contribution to sustainable technological practices.
This section further analyses the impact of traffic patterns and ABR segment size on the Ethernet NIC energy consumption.

\paragraph*{Traffic pattern}
La Oliva~\etal~\cite{DELAOLIVA2013599} evaluated the performance of the EEE standard under different traffic conditions for UDP-based video streaming: \begin{enumerate*} \item Poisson, transmitting Ethernet traffic following the named distribution, \item real, simulating the back-to-back transmission nature of video streaming, and \item group of pictures (GoP), aggregating the whole data before transmission.\end{enumerate*}
The results showed a larger back-to-back transmission of video streams per wake-up and sleep-down cycle, reducing the power-mode transitions and minimizing the energy overhead caused by switching between modes. The study also showed that the scalable AVC codec generated GoP-size data bursts and was more energy efficient than the non-scalable AVC codec.

\paragraph*{ABR segment sizes}
Vargas~\etal~\cite{vargas2023contribution} investigated video traffic patterns within IPTV and DASH streaming applications and their relationship with the Ethernet traffic. The results suggest that larger segment sizes significantly enhance energy savings given a specific bandwidth, reaching approximately \qty{50}{\percent} when transitioning from \qty{4}{\second} to \qty{10}{\second} segments. Therefore, EEE integration appears promising for servers and clients involved in DASH traffic, regardless of the segment duration.
 % segment efficiency correlates with increased bandwidth and size. In particular, larger segment sizes for a given bandwidth substantially enhance segment efficiency. 

\subsubsection{Network types}
Studies showed that WiFi NICs are more energy efficient than LTE for video streaming. For instance, Huang~\etal~\cite{huang2012close} concluded that, despite several new power-saving improvements, LTE is \num{23} times less power-efficient than WiFi. In some experiments, LTE is even \num{1.72} times less efficient than 3G for small data transfers (\ie one packet) due to a prolonged high-power tail. However, LTE is more energy efficient than 3G when fully utilized. For example,  3G requires \num{21.5} times the energy of LTE and \num{34.77} times the energy of WiFi for downloading \qty{10}{\mega\byte} of data. Zhang~\etal~\cite{zhang2016profiling} observed that the overall NIC power consumption under LTE networks is higher than in WiFi for regular and \SI{360}{\degree} videos.

\begin{table}[!t]
\caption {Related work classification of decoding on the end-user devices.}
\label{tbl:related:decod}
\vspace*{-0.3cm}
\resizebox{\textwidth}{!}{
\begin{tabular}{|l|l|l|l|l|l|l|l|l|}
\hline

\multirow{2}{*}{\emph{Work}} & \multirow{2}{*}{\emph{Goal / Method}} & \multicolumn{4}{l|}{\emph{Encoding parameters}} &\multicolumn{2}{l|}{\emph{Machine}} & \multirow{1}{*}{\emph{Energy}}\\
\cline{3-6}
\cline{7-8}
& & \emph{CD} & \emph{RS}& \emph{FR [fps]} &  \emph{QP} & \emph{Type} & \emph{Brand} &\emph{tool} \\
\hline\hline

\multirow{2}{*}{\cite{katsenou_energy-rate-quality_2022}} & Software energy  & AV1 (SVT-AV1), VVC (VVdeC) & \SI{416}{\pixel}, \SI{832}{\pixel},   &\num{20}, \num{30}, & \num{22}, \num{27}, & \multirow{2}{*}{PC} & \multirow{2}{*}{Intel CPU} & \multirow{2}{*}{RAPL} \\
&analysis&  VP9 (VP9), HEVC (FFmpeg) &\SI{1920}{\pixel} & \num{50}, \num{60}&\num{32}, \num{37}& && \\
\hline

\multirow{2}{*}{\cite{kranzler_comparative_2020}} & Software energy  & \multirow{2}{*}{HEVC (HM), VVC (VTM)} & \SI{416}{\pixel}, \SI{832}{\pixel}, & \num{20}, \num{30}, & \num{22}, \num{27},    & \multirow{2}{*}{PC }& \multirow{2}{*}{Intel CPU} & \multirow{2}{*}{RAPL}\\
&analysis&&\SI{1920}{\pixel}, \SI{3840}{\pixel} & \num{50}, \num{60} &  \num{32}, \num{37}   &&& \\
\hline

\multirow{2}{*}{\cite{yue2020energy}} & Linear hardware  &  \multirow{2}{*}{AVC, HEVC} & \SI{144}{\pixel}, \SI{240}{\pixel}, \SI{360}{\pixel}, & \num{7}, \num{15}, &  \multirow{2}{*}{--}
 & \multirow{2}{*}{Smartphone} %LG V20, Moto G5  
 &Qualcomm & Monsoon power \\
&power model&&\SI{480}{\pixel}, \SI{720}{\pixel},  \SI{1080}{\pixel} & \num{30}, \num{60}&&&Snapdragon& monitor \cite{monsoonpowermonitor}\\
\hline

\multirow{2}{*}{\cite{turkkan2022greenabr}} & \multirow{1}{*}{AI-based} & \multirow{2}{*}{AVC} & $320\times180$ -  &\multirow{2}{*}{\num{24}} & \multirow{2}{*}{--}& \multirow{2}{*}{Smartphone} & \multirow{1}{*}{Qualcomm}  & 
Monsoon power  \\
&playback model &&\SI{2160}{\pixel} &&&&Snapdragon & monitor \cite{monsoonpowermonitor} \\
\hline

\multirow{3}{*}{\cite{herglotz2022modeling}} & Linear hardware and  & AVC (Direct3D 11 hardware  & \multirow{3}{*}{\SI{416}{\pixel}-\SI{1920}{\pixel}} & \multirow{3}{*}{\num{24} - \num{60}}&\multirow{3}{*}{--}&\multirow{3}{*}{PC, Laptop} & \multirow{3}{*}{Intel CPU} & ZES Zimmer  \\
& software power model & acceleration, FFmpeg), &&&&&&LMG95 \\
 &    & VP9 (FFmpeg) &&& &&& \\
\hline

\multirow{2}{*}{\cite{herglotz2020power}} & Linear hardware   & \multirow{2}{*}{AVC, HEVC} & \SI{416}{\pixel}, \SI{832}{\pixel},  & \num{20}, \num{24}, \num{25},&\multirow{2}{*}{--} & \multirow{2}{*}{Smartphone} & Qualcomm  & \multirow{2}{*}{Power meter} \\ 
& power model & \multirow{2}{*}{--}& \SI{1280}{\pixel}, \SI{1920}{\pixel} &\num{30}, \num{50}, \num{60}&&&Snapdragon& \\
\hline

\multirow{2}{*}{\cite{herglotz2013modeling} }& Heuristic software  & \multirow{2}{*}{HEVC (HM)} & \multirow{2}{*}{\SI{416}{\pixel}, \SI{832}{\pixel}, \SI{1920}{\pixel}} & \multirow{2}{*}{--}&\num{10}, \num{32},  & \multirow{2}{*}{SBC} & \multirow{2}{*}{ARM CPU} & \multirow{2}{*}{Agilent 34401A} \\
& energy model &&&& \num{45} &&& \\ 
\hline

\multirow{3}{*}{\cite{herglotz2014modeling}} & Heuristic software  & \multirow{3}{*}{HEVC (HM)} & \multirow{3}{*}{\SI{416}{\pixel}, \SI{832}{\pixel}, \SI{1920}{\pixel}} &\multirow{3}{*}{--}&\num{2}, \num{4}, \num{8}, & \multirow{3}{*}{SBC} & \multirow{3}{*}{--} & \multirow{3}{*}{Agilent 34401A} \\
& energy model &&&&  \num{10}, \num{12}, &&& \\
&  &&&&  \num{32}, \num{45}&&& \\
\hline

\multirow{2}{*}{\cite{herglotz2016modeling}} & Heuristic hardware and  & \multirow{2}{*}{HEVC (HM, FFmpeg)} & \SI{416}{\pixel}, \SI{832}{\pixel}, &\multirow{2}{*}{--}& \num{10}, \num{32},  & \multirow{2}{*}{SBC} & \multirow{2}{*}{Intel CPU} &  ZES Zimmer’s\\
& software energy model && \SI{1280}{\pixel}, \SI{1920}{\pixel}, \SI{2560}{\pixel}&&\num{45}&&&  LMG95  \\
\hline

\multirow{2}{*}{\cite{ren2014energy}}& Hardware energy-aware &\multirow{2}{*}{AVC, HEVC} & \multirow{2}{*}{\SI{144}{\pixel}, \SI{288}{\pixel}, \SI{720}{\pixel}} & \multirow{2}{*}{--} &\multirow{2}{*}{--}& \multirow{2}{*}{SBC} &\multirow{2}{*}{--} & Performance monitor  \\
& model&&&&&&& unit, PAPI~\cite{PAPI}\\

\hline

\multirow{2}{*}{\cite{herglotz2015estimating}} & Linear regression  & \multirow{2}{*}{HEVC (HM)} & \SI{416}{\pixel}, \SI{832}{\pixel}& \multirow{2}{*}{\num{16}, \num{30}, \num{40}}& \num{10}, \num{32},  & \multirow{2}{*}{SBC} & \multirow{2}{*}{ARM CPU} & \multirow{2}{*}{LMG95} \\
& software model && \SI{1280}{\pixel}, \SI{1920}{\pixel}, \SI{2560}{\pixel} &&\num{45} &&& \\
\hline

\multirow{2}{*}{\cite{herglotz_decoding-energy-rate-distortion_2019}} & Heuristic software  & \multirow{2}{*}{HEVC (HM, FFmpeg)} &\SI{416}{\pixel}, \SI{832}{\pixel}, & \multirow{2}{*}{\num{24} - \num{60}}& \num{22}, \num{27},  & \multirow{2}{*}{SBC} &\multirow{2}{*}{--} & ZES Zimmer's  \\
& energy model  &&\SI{1920}{\pixel}, \SI{2560}{\pixel}&&\num{32}, \num{37}&&& LMG95\\
\hline

\multirow{2}{*}{\cite{kranzler2022optimized}} & Software energy  & \multirow{2}{*}{VVC (VVdeC)} & \multirow{2}{*}{\SI{720}{\pixel}, \SI{1080}{\pixel}} & \multirow{2}{*}{\num{24}, \num{30}, \num{60}} & \multirow{2}{*}{\num{32}, \num{37}}  & \multirow{2}{*}{PC} & \multirow{2}{*}{Intel CPU} & \multirow{2}{*}{RAPL} \\
 & analysis &  & & &  &   & &  \\

\hline 

\end{tabular}
}
\end{table}

% %%%%%%%% TODO: add this work: HEVC hardware vs software decoding: An objective energy consumption analysis and comparison  2021

\subsection{Video decoding}
With the widespread creation of video coding standards, the complexity of new algorithms has increased, which has also increased the decoding time and energy consumption~\cite{herglotz2015estimating}.

\subsubsection{Video decoding energy analysis}\label{sec:dec:anal}
Several factors influence video decoding energy consumption, categorized into four groups: \begin{enumerate*} \item decoding mode, \item codec, \item encoding configuration, and \item decoding configuration\end{enumerate*}, explained in this section.

\paragraph{Decoding mode}
Khernache~\etal~\cite{bey_ahmed_khernache_hevc_2021} investigated the power consumption of hardware and software decoding on mobile platforms.

\paragraph*{Software decoding} It consumes, on average, less than \SI{50}{\percent} of the total power consumption of the tested platforms (smartphones).  However, scaling up the video quality parameters significantly increases energy consumption. For instance, scaling the resolution from \qtyrange{720}{2160}{\pixel} consumes \num{13.76} times more energy;
\paragraph*{Hardware decoding} It consumes in comparison less than \SI{30}{\percent} of the total power consumption of the tested platforms. The impact of bitrate, framerate, and resolution is minimal, and scaling the resolution from \qtyrange{720}{2160}{\pixel} consumes only \num{2.14} times more energy. Hardware decoding uses one-fourth less energy than software decoding for mobile devices with \qty{1080}{\pixel} or lower resolution, and its advantage increases with higher resolutions or framerates. 

\paragraph{Codec}
The codec significantly impacts the device power consumption~\cite{herglotz2022modeling}. For example, the average power consumed for a streaming session on a laptop is \qty{13.5}{\watt} for AVC, \qty{17.6}{\watt} for VP9. For PCs, it is \qty{81.3}{\watt}, respectively \qty{76.3}{\watt}. 
Katsenou~\etal~\cite{katsenou_energy-rate-quality_2022} examined the impact of decoding codecs (\ie AV1, VP9, VVC, HEVC) on energy consumption and video quality, aiming to identify the most efficient balance. 
%The findings indicate that, out of the codecs analyzed in the research (AV1, VP9, VVC, HEVC), AV1 offers the best compromise between energy consumption and quality. 
While AV1 and VVC show similar trends in the decoding energy-bitrate curve, AV1 needs less decoding energy. The authors in~\cite{viitanen_hevc_2022} found that HEVC has up to \SI{87}{\percent} more decoding complexity than its predecessor AVC in various settings, motion compensation and loop filtering being the main contributors. The work in~\cite{kranzler_comparative_2020} revealed that the decoding energy for VVC is higher than HEVC, depending on the coding configurations. For random access configuration, for example, the decoding energy increases by over \qty{80}{\percent} and the decoding time by over \qty{70}{\percent}.
 
\paragraph{Encoding configuration} This factor comprises resolution, framerate, and preset.

\paragraph*{Resolution} Energy consumption increases with the higher resolution %with larger higher
and decoding overhead. Furthermore, the decoding power, defined as the total power subtracted from the idle power of the CPU, screen, and network, is not sensitive to the video content and bitrate variations within the same resolution~\cite{yue2020energy}. %In this work.

\paragraph*{Framerate} Yue~\etal~\cite{yue2020energy}
%studied the impact of framerate on decoding energy consumption and
discovered that increasing the framerate of a video streaming from \numrange{30}{60} frames per second for a given mobile phone resolution doubles the energy decoding. For other cases,  increasing from \numrange{7}{15} and \numrange{15}{30} the increase was more than double. 

\paragraph*{Preset} Herglotz~\etal~\cite{herglotz_decoding-energy_2020} studied various presets and tuning options for the x265 encoder resulting in the lowest decoding energy. A decoding-energy-rate-distortion (DERD) algorithm found the \texttt{medium} preset as optimal with \texttt{fastdecode} tuning and improved DERD optimization. The results showed over \qty{25}{\percent} less energy for the OpenHEVC and HM software decoders with a minor \qty{0.39}{\percent} runtime reduction and an average Bjøntegaard-Delta rate loss of \qty{38.2}{\percent},

\paragraph{Decoding configuration}
Optimized decoding algorithms can substantially reduce energy expenditure.
The work in \cite{kranzler_comparative_2020}  revealed a notable decrease in decoding time of Single Instruction Multiple Data (SIMD) on the VVC decoding using the VTM decoder~\cite{VVCSoftware}, consequently reducing energy consumption. For random access configuration, for instance, disabling the SIMD in the VTM decoder produces a \qty{211.76}{\percent} increase in decoding time and \qty{207.04}{\percent} increase in energy consumption compared to the HM decoder. Furthermore, an evaluation of various VVC codec tools (\ie matrix-weighted intra-prediction, adaptive motion vector resolution, triangular partition mode, low-frequency non-separable transform, multiple transform set) revealed %enhanced energy efficiency.
%The authors examined the effect of each tool by switching them off one by one and measuring the decoding time and energy consumption.
a \qty{17}{\percent} lower average decoding energy based on a proposed coding configuration.

\subsubsection{Energy-aware video decoding}\label{sec:dec:ener-aware}
Several works focused on proposing models to estimate decoding energy consumption, as presented in Table~\ref{tbl:related:decod}.  

\paragraph*{GreenABR} Turkkan~\etal~\cite{turkkan2022greenabr} introduced GreenABR, a deep reinforcement learning method to optimize ABR decoding energy consumption that dynamically adapts to network conditions without prior information for enhanced user QoE. GreenABR achieved \qty{57}{\percent} reduction in energy consumption and \qty{22}{\percent} increase in QoE compared to Pensieve~\cite{mao2017neural} and Bola~\cite{spiteri2020bola} methods.

\paragraph*{Power estimation models} Herglotz~\etal~\cite{herglotz2022modeling} measured power and quality using a crowdsourced dataset~\cite{robitza2020you} of \num{447000} streaming events and proposed a linear power model for laptops and PCs. They optimized decoding power consumption and QoE by tuning the streaming parameters, such as video codec, resolution, and bitrate. Another power model for mobile devices~\cite{herglotz2020power} uses parameters like framerate, pixels per frame, bitrate, and output pixel rendering, achieving low estimation errors below \qty{7.61}{\percent}. Moreover, they provided energy models for HEVC based on bitstream features~\cite{herglotz2013modeling, herglotz2014modeling, herglotz2016modeling} for intra-coded videos, videos with in-loop filters, and videos with different decoding solutions, with errors ranging from \qty{2.34}{\percent} to \qty{15}{\percent}.

\paragraph*{Extended rate-distortion optimization} The work in \cite{herglotz_decoding-energy-rate-distortion_2019} extends the standard rate-distortion optimization with decoding energy estimated by the encoder using a feature-based energy model. The encoder minimizes the DERD costs and produces bitstreams that require less decoding energy. Tests with HEVC decoders show saving up to \qty{30}{\percent} on decoding energy while keeping the visual quality at the cost of \qtyrange{20}{50}{\percent} more bitrate. The work in \cite{kranzler2022optimized} addresses the VVC decoding energy consumption by configuring a set of coding tools and parameters, such as depth of partitioning and affine motion estimation, to optimize decoding energy by \qty{34.07}{\percent}.

\paragraph*{Green metadata} It is a standard~\cite{greenmpeg,fernandes2015} for energy-efficient video consumption for tasks such as video decoding or displaying using metadata to inform the user device about the quantity of unpacking loaded for the data. The metadata allows the device to anticipate the energy consumption during decoding and adjust its processors to an optimized power state. On the display side, enhancing the gamma (i.e., brightness levels) in the picture and lowering the brightness level used for illumination can reduce the power consumption by \qty{30}{\percent} and maintain a high QoE.

\begin{wrapfigure}{R}{0pt}
    \centering
    \includegraphics[width=0.4\textwidth]{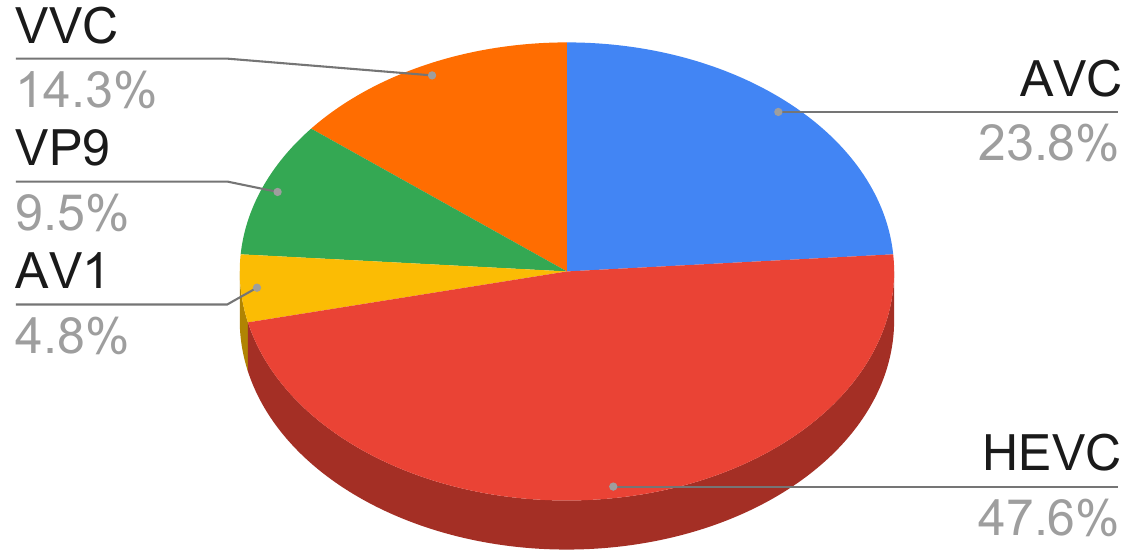} 
    \caption{\centering Distribution of codecs for the related video decoding works for  in Table~\ref{tbl:related:decod}.}
    %\vspace{-5pt}
    \label{fig:Dec_codec} 
\end{wrapfigure}

\subsubsection{Video decoding codecs distribution}
Table~\ref{tbl:related:decod} shows various study efforts on the energy consumption for different video decoding techniques (see Sections~\ref{sec:dec:anal} and~\ref{sec:dec:ener-aware}). As observed in Figure~\ref{fig:Dec_codec}, most research works investigate AVC~\cite{AVC} and HEVC~\cite{sullivan2012overview} standards. There are also several investigations on the energy consumption analysis of VVC~\cite{bross_overview_2021}, but limited studies on AV1~\cite{han_technical_2021} and VP9 codecs (see Figure~\ref{fig:Dec_codec}).

%However, there is still room for energy consumption analysis in other codecs, such as VP9, AV1~\cite{han_technical_2021}, and VVC~\cite{bross_overview_2021} for more investigation. In Figure~\ref{fig:Dec_codec}, the number of times each codec has been analyzed with the purpose of investigating the decoding energy consumption in this survey (based on Table~\ref{tbl:related:decod}) is illustrated. As depicted in this figure, about \SI{80}{\percent} of the chart belongs to AVC and HEVC, while the rest is divided among AV1, VP9, and VVC. 
%For further details, refer to Figure~\ref{fig:Dec_codec}.

\begin{table*}[!t]
\small
\centering
\footnotesize
\caption {Related work classification of display on the end-user devices.}
\label{tbl:related:disp}
\vspace*{-0.3cm}
\resizebox{\textwidth}{!}{
\begin{tabular}{|l|l|l|l|l|}
\hline
\multirow{2}{*}{\emph{Work}} & \multirow{2}{*}{\emph{Goal / Method}} &  \multicolumn{2}{l|}{\emph{Machine}} & \multirow{2}{*}{\emph{Energy tool}}\\
\cline{3-4}
&    & \emph{Type} & \emph{Brand} & \\
\hline\hline

\cite{yue2020energy} & Linear LCD power model & Smartphone &  LG V20, Moto G5 & Monsoon power monitor \cite{monsoonpowermonitor} \\

\hline

\cite{herglotz2020power} & Linear LCD power model & Smartphone & Fairphone 2, LG H812 & Power meter \\

\hline

\multirow{2}{*}{\cite{wan2017detecting}} & \multirow{2}{*}{Analysis of OLED energy model} & \multirow{2}{*}{Smartphone} & Samsung Galaxy S2,  & \multirow{2}{*}{Monsoon power monitor \cite{monsoonpowermonitor}} \\ 
&&&S5, Nexus & \\
%\makecell{~\cite{wan2017detecting}} & \makecell{Analysis of OLED display \\energy model to detect \\energy hotspots  } & \makecell{ Heuristic} 
% & \makecell{Real exper.: \\Smartphones: \\ Samsung Galaxy S2,\\ S5, and Nexus } & \makecell{Energy: HW. \\Monsoon \\power monitor\\ \cite{monsoonpowermonitor}}\\
\hline

\multirow{2}{*}{\cite{dong2011power}} & \multirow{2}{*}{Heuristic OLED power model} & Monitor,  &QVGA OLED,   & DAQ Board Measurement  \\
&&Smartphone& Nokia N85 &Computing  \cite{DAQMC} \\

%\makecell{~\cite{dong2011power}} & \makecell{Dev. OLED display power model  } & \makecell{ Heuristic} & \makecell{Real exper.: \\QVGA OLED display module, \&\\ Nokia N85 smartphone } & \makecell{Current: HW. \\DAQ Board \\(Measurement \\Computing  \cite{DAQMC})}\\
\hline

\cite{gui2016lightweight} & Heuristic power model & Smartphone & Samsung Galaxy S2 & Monsoon power monitor \cite{monsoonpowermonitor}  \\ 
%\makecell{~\cite{gui2016lightweight}} & \makecell{Dev. display power model \\ to estimate mobile and \\energy consumption } & \makecell{ Heuristic}  & \makecell{Real exper.: \\Samsung Galaxy S2 } & \makecell{Energy: HW. \\Monsoon \\power monitor\\ \cite{monsoonpowermonitor}}\\
\hline

\multirow{2}{*}{\cite{mittal2012empowering}} & \multirow{2}{*}{Heuristic AMOLED power model} & \multirow{2}{*}{Smartphone} &Samsung Focus, HTC HD7, & \multirow{2}{*}{WattsOn \cite{mittal2012empowering}} \\
&&&HTC Trophy, HTC Arrive& \\

%\makecell{~\cite{mittal2012empowering}} & \makecell{Dev. AMOLED display \\ power model \& Dev. energy \\emulation tool  } & \makecell{ Heuristic}  & \makecell{Real exper.: \\Smartphones: \\ Samsung Focus, \\ HTC HD7, \\ HTC Trophy, \\ HTC Arrive} & \makecell{Energy. SW.: \\WattsOn \cite{mittal2012empowering}}\\
\hline

\multirow{2}{*}{\cite{demarty2023display}} & \multirow{2}{*}{DL-based power model} & \multirow{2}{*}{Monitor} & \multirow{2}{*}{Sony 55AF9, LG 42C2}& Voltcraft energy logger  \\
&&& &4000F wattmeter\\
%\makecell{~\cite{demarty2023display}} & \makecell{Dev. display power model } & \makecell{DL}  & \makecell{Real exper.: \\Screens: \\ Sony 55AF9,\\ LG 42C2} & \makecell{Energy. HW.: \\Voltcraft energy \\logger 4000F\\ wattmeter }\\
\hline

% \specialrule{.12em}{.05em}{.05em}
% \specialrule{.12em}{.05em}{.05em}

% a &   c & e & f & g & f & i\\
% \hline

\end{tabular}}

\end{table*}

\subsection{Display}\label{ssec:display}
%The screen power mainly depends on its brightness level.

The fraction of energy consumed by the display can be more significant than the other components since it is constantly active throughout an application's runtime. Thus, an accurate estimate is essential for this dominant energy consumer.
The power model of the display depends on its technology. For LCDs, the power consumption depends mainly on the brightness level of the screen~\cite{wan2017detecting}. On the other hand, the energy consumption of more modern technologies such as Organic Light-Emitting Diode (OLED) and Active Matrix OLED (AMOLED) depends on the red, green, and blue (RGB) elements of its pixels with varying levels of luminance efficiency~\cite{dong2011power}.
%Consequently, the power consumption of a pixel is directly influenced by the colors on the display

%Developing power models for different display technologies has been studied in several research. 

\paragraph*{LCD technology}
The works in \cite{yue2020energy, herglotz2020power} model the LCD power consumption using the backlight brightness. Yue~\etal~\cite{yue2020energy} determine the screen power consumption by measuring three different brightness levels on two phones in two settings: \begin{enumerate*} \item both screen and CPU turned on, and \item CPU on and screen off\end{enumerate*}.

\begin{description}[font=\normalfont\itshape]
    \item[LG phone] has a screen average power consumption of \qty{293}{\milli\watt}, \qty{440}{\milli\watt}, and \qty{754}{\milli\watt} for brightness levels of \qty{30}{\percent}, \qty{50}{\percent}, and \qty{80}{\percent}, respectively, with coefficient of variation below \num{0.01}.
    \item[Moto phone] has power consumption of \qty{446}{\milli\watt}, \qty{573}{\milli\watt}, and \qty{858}{\milli\watt}, respectively.
\end{description}

\paragraph*{OLED and AMOLED technology}
For OLED and AMOLED, researchers~\cite{gui2016lightweight, wan2017detecting, dong2011power} involve \num{16} screenshots of different graphical user interfaces displayed from the application runtime for each color component (\ie red, green, blue) at varying intensities while keeping the other two components at zero. The process first measures the power consumption of a completely black screen to establish a baseline and then the power used to display each screenshot in the set. Afterward, it obtains the power consumed by the RGB components by subtracting the black screen power. Finally, it applies gamma correction to the RGB values, followed by linear regression to determine the coefficients for each component. Dong~\etal~\cite{dong2009power} propose a power model to estimate the power consumption at three levels:

\begin{description}[font=\normalfont\itshape]
    \item[Pixel-level model] obtains an accuracy of \qty{99}{\percent} in power estimation;
    \item[Image-level model] can reduce the computation cost with a \qty{90}{\percent} accuracy by gathering the power of small groups of pixels;
    \item[Code-level model] reaches \qty{95}{\percent} power forecast accuracy using the graphical interface specification.% objects. 
\end{description}

The work in~\cite{mittal2012empowering} provides a tool for developers to estimate the energy usage of their mobile application and an energy model for AMOLED displays.

\paragraph*{Technology-agnostic}
Recently, Demarty~\etal~\cite{demarty2023display} proposed a deep learning model to predict the power consumption of displaying an image on a specific screen, agnostic to the implementation technology. Le Meur~\etal~\cite{le2023deep} proposed a lightweight deep model to maintain the QoE of an image and reduce energy consumption with an energy savings rate. They obtained a dimming map to adjust the image's luminance based on the information learned by the neural network during its training according to the energy-saving rate.

% \subsection{Mobile devices}
% Suski~\etal~\cite{suski2020all} surveyed that the \emph{global warming potential} (\emph{GWP}) and the intensity of video streaming highly depends on the user's requirement and their device types. The results show that the largest \emph{GWP} for streaming on laptops and smart TVs stems from production and operational electricity demands. Moreover, the survey reveals that smartphones have ten times more environmental impacts than smart TVs.

% https://dl.acm.org/doi/pdf/10.1145/3339825.3391867?casa_token=gtEUs6HD4v4AAAAA:2EVMWfHKZwG8jlqQ7zdVigy_Qme3w9XzBwcW_-j5n31HBI-ZCKH0rWYWrAbK5kCB_bpBVaxjNwq8MA s [35, 54, 72].

% \subsubsection{Experiment design and measurement tool}
% Table~\ref{tbl:related:disp} shows that experimental evaluation is mainly dominated by smartphones. Energy measurements are mostly taken by a power meter, such as the Monsoon power monitor, primarily for accurate energy measurement.

\subsection{Research gap}
\label{sec:rg_dec}
\paragraph*{Optimized video playback}
%Designing a video player that integrates various video segment sizes. For instance, in optimal network conditions, the player can offer the option of larger segments to the client.
A video player that adaptively handles various segment sizes can offer viewers the option of larger video segments. For example, Schwarzmann~\etal~\cite{Schwarzmann2020} investigated variable segment sizes but neglected energy consumption. Additionally, optimizing the video player for larger buffers allows fetching multiple segments in parallel that for inconsistent user behavior is consistent. These improvements can achieve a good balance between high QoE and energy saving.

\paragraph*{Optimized encoder parameters} Appropriate selection of power-aware encoder parameters can significantly affect the decoding energy and improve the models' comprehensiveness.
%Considering the encoding parameters for modeling the decoding energy can improve the models' comprehensiveness in achieving an energy-efficient decoding procedure.

\paragraph*{Technology-agnostic display energy models}
Creating precise energy models independent of specific technologies is essential to accurately estimate the energy consumption of devices such as smartphones and various displays. This necessity arises from the advent of emerging display technologies such as Quantum dot LED (QLED).
\section{Datasets and tools}
\label{sec:dataset&tools}

According to the literature studied in this work, various datasets and tools exist for real or simulated experiments, summarized in Tables~\ref{tbl:related:dataset} and~\ref{tbl:related:tools}.

Table~\ref{tbl:related:dataset} covers the energy consumption ranges for the video applications running on the end-user's devices or cloud data centers and the calculation of the carbon footprint based on the solar power plant available in specific regions.  Moreover, Electricity Maps presents online visualization data on the amount of power produced by different sources~\cite{gmsys2023afzal}.

Table~\ref{tbl:related:tools} presents existing measurement tools for energy, power,  or \ch{CO2} emissions. The tools investigated the support of various video encoding and decoding applications, from containerized to parallel ones executed on multicore machines. We categorized the software-based tools into hardware-agnostic and model-based tools. Moreover, Intel, HP, and Nvidia-specific products provide a measurement capability for power and energy consumption, which helps eco-friendly customers monitor and estimate the impact of their video streams on the climate.
Additionally, several power metering tools can collect measurements related to physical device consumption and monitor the results for user applications.

\begin{table}[!t]
\caption{Energy consumption and carbon emissions datasets.}
\label{tbl:related:dataset}
\resizebox{\textwidth}{!}{
\begin{tabular}{|l|l|}
\hline
 \emph{Dataset} & \emph{Description} \\
\hline\hline
GreenABR \cite{greenabr} &  \begin{tabular}[c]{@{}l@{}}Local playback energy consumption of videos in several genres on a Samsung Galaxy S4 mobile device\end{tabular}   \\ \hline
\begin{tabular}[c]{@{}l@{}}Green500 \cite{green500} \end{tabular}& Power consumption of \num{500} energy efficient machines \\ \hline
 Solar energy traces~\cite{li2015opportunistic} & Solar power data records of every hour (\ie \num{24} values per day) provided by the University of Nantes, France
\\\hline
SPECpower~\cite{specpower2014}& \begin{tabular}[c]{@{}l@{}}Servers' power and  performance characteristics  measurements \end{tabular}
\\\hline

Electricity Maps~\cite{tranberg2019real,elecmap} & \begin{tabular}[c]{@{}l@{}}Online visualization of electricity and \ch{CO2} dataset on an hourly basis across over \num{160} regions\end{tabular}
\\\hline

% &Robitza~\etal~\cite{robitza2020you}& \makecell{400,000 video playbacks \\from more than 2,000 users}
% \\\cline{2-3}

Software encoder energy~\cite{chachou2023energy2} & \begin{tabular}[c]{@{}l@{}}Energy consumption and carbon emissions of x264, x265, libvpx-vp9, VVenC, and SVT-AV1 encoders\end{tabular}
\\\hline
\end{tabular}}
\end{table}

\begin{table}[!t]
\begin{center}
%\footnotesize
\centering
\caption{Tools classification for energy consumption and carbon emission measurements.}
\label{tbl:related:tools}
\resizebox{\textwidth}{!}{
\begin{tabular}{|c|l|l|} % Changed the last column to X type for better text wrapping
%\midrule  % Used \toprule instead of \specialrule for better spacing
%\midrule 
\cline{2-3}
\multicolumn{1}{c|}{}&\emph{{Tool name}} & \emph{{Description}} \\
\midrule % Used \midrule instead of \specialrule for better spacing
\midrule 
\multirow{10}{.7cm}{\centering \textit{\rotatebox[origin=c]{90}{Non hardware-customized}}}
&\multirow{1}{.7cm}{PowerAPI~\cite{%bourdon2013powerapi,
powerapi}}& Software-defined power meter %to measure the power consumption
of an application
\\\cline{2-3}
&\multirow{1}{*}{Turbostat~\cite{turbostat}}
%,estim-ec2-power
& Frequency, temperature, %idle power state, 
and power on x86 processors
\\\cline{2-3}

&\multirow{1}{*}{Climatiq~\cite{climatiq}}& GHG emission calculations
%Power meter for data centers based on the CPU, memory, and storage that services use, the utilization of the electricity, and the geographical location of the data center 
\\\cline{2-3}
&\multirow{1}{*}{Nornir~\cite{de2018simplifying}} & Power meter for parallel applications on shared memory multicore machines
\\\cline{2-3}

&\multirow{1}{*}{PowerTutor~\cite{powertutor}} & Online power estimation of CPU, NIC and display %tool implemented for Android 
%, and GPS receiver 
\\\cline{2-3}
&\multirow{1}{*}{PAPI~\cite{PAPI}} & Performance counters including CPUs, GPUs, accelerators, interconnects, I/O systems, power interfaces
%, and virtual cloud environments
\\
%To gather performance counter data including various CPUs, GPUs, accelerators, interconnects, I/O systems, power interfaces, and virtual cloud environments\\
\cline{2-3}
&\multirow{1}{*}{CodeCarbon \cite{codecarbon}} &
Device energy consumption and \ch{CO2} emissions using RAPL for Intel CPU and PyNVML for Nvidia GPU.\\ \cline{2-3}

&\multirow{1}{*}{Greenhouse Gas %Equivalencies 
Calc.~\cite{usa-epa-ghg}}&
Converter of energy data to equivalent \ch{CO2} car or power plant emissions %into concrete terms, such as %the annual
\\ \cline{2-3}
&\multirow{1}{*}{Carbonalyser~\cite{carbonanalyzer}} & 
Device’s electricity consumption as Firefox add-on~\cite{ShiftProjectReport2020} \\\cline{2-3}
&Cloud carbon footprint~\cite{cloudcarbonfootprint}& Carbon emission calculator for cloud service providers
\\
\midrule % Used \midrule instead of \specialrule for better spacing
\midrule 
\multirow{5}{.7cm}{\centering \textit{\rotatebox[origin=c]{90}{\makecell{Model-based\\custom}}}}
&DockerCap~\cite{asnaghi2016dockercap} &Power-aware orchestrator for a cluster of devices
\\\cline{2-3}
&\multirow{1}{*}{DEEP-mon~\cite{brondolin2018deep}} &Power monitoring for Linux  and containerized applications based on RAPL and DVFS 
\\\cline{2-3}
&HEATS~\cite{rocha2019heats} &Kubernetes and cAdvisor-based energy-aware scheduler\\
\cline{2-3}
&\multirow{1}{*}{DIMPACT~\cite{dimpact}}& %Developed as part of the "responsible media forum" \cite{responsiblemedia} in 2015 by media, entertainment and technology companies (now involving more than 20 companies such as Netflix, Google, Spotify, etc.)  
Energy consumption and \ch{CO2} emissions estimator for digital media products and services
\\\cline{2-3}

&\multirow{1}{*}{StreamingCalc~\cite{makonin2022calculating}}& Carbon footprint calculator for streaming media %with an end-to-end model
\\ 
\midrule % Used \midrule instead of \specialrule for better spacing
\midrule 

\multirow{4}{.7cm}{\centering \rotatebox[origin=c]{90}{\makecell{\textit{Hardware-}\\\textit{customized}}}} &\multirow{1}{*}{Intel$^\circledR$ PCM~\cite{intelpcm}}&  Energy consumption, CPU and memory utilization measurement%The tool considers the machine-specific registers (MSR) using RAPL counters to disclose the energy consumption of the application execution.
\\
%\arrayrulecolor{black}
\cline{2-3}

&RAPL~\cite{khan2018rapl} &\texttt{Intel$^\circledR$} processor power meter %The research conducted by \\Fahad~\etal~\cite{fahad2019comparative} shows a strong correlation between RAPL readings and system power meters.
\\\cline{2-3}
&\texttt{HP$^\circledR$} CACTI \cite{cacti, cacti2022cacti}
%~\cite{cacti2022cacti}
& Dynamic power along with the memory utilization for HP machines
\\\cline{2-3}
&pyNVML~\cite{nmvl}& 
Nvidia$^\circledR$ GPU Power consumption management and monitoring
\\
\midrule % Used \midrule instead of \specialrule for better spacing
\midrule 
\multirow{1}{*}[+0.6em]{\rotatebox[origin=l]{90}{\makecell{\textit{Hard}-\\\textit{ware}}}} &
\multicolumn{2}{l|}{\makecell{Watts Up? Pro, Pro ES~\cite{ramseyer2013watts}, TechPowerUp GPU-Z \cite{techpowerup}, Monsoon \cite{monsoonpowermonitor}, Agilent 34401A, ZES Zimmer's LMG95, \\ Voltcraft Energy Logger 4000F, Eaton Managed ePDU \cite{eaton}}}
    % \begin{itemize}[left=0.2cm]
    %     \item Watts Up? Pro, Pro ES~\cite{ramseyer2013watts}
    %     \item TechPowerUp GPU-Z \cite{techpowerup}
    %     \item Monsoon \cite{monsoonpowermonitor}  
    %     \item Agilent 34401A 
    %     \item ZES Zimmer's LMG95
    %     \item Voltcraft Energy Logger 4000F 
    %     \item Eaton Managed ePDU \cite{eaton}
    % \end{itemize}
    % &   
    % \multirow{7}{*}{Power measuring devices}
\\
%\arrayrulecolor{black}
%\cline{2-3}
\bottomrule % Used \bottomrule instead of \specialrule for better spacing
\bottomrule
\end{tabular}}
\end{center}
\end{table}
\section{Open research issues}
\label{sec:openResearches}
Section~\ref{sec:rg_enc} highlighted research gaps on content provisioning, while Section~\ref{sec:rg_dec} discussed gaps on content consumption. This section provides a comprehensive perspective on the video streaming workflow concerning open research issues. 
%by identifying research issues that consider the entire process.
%of video encoding, storage, retrieval, decoding, and display.

\paragraph*{Holistic energy-efficient system design}
Video streaming systems involve a complex network of interconnected components, crucial to optimize individual energy consumption to prevent unnecessary processing burdens~\cite{koziri2018efficient}. For instance, advanced video codecs can effectively reduce data transmission and its energy requirements but can also introduce higher encoding and decoding complexities, demanding additional resources. Therefore, a comprehensive system design to maximize energy savings is more advantageous than solely optimizing individual components~\cite{herglotz2022modeling}.

\paragraph*{In-depth internal measurements and analysis} Many current research works measure energy consumption per \unit{\giga\byte} or \unit{\hour} of video streaming, and ignore crucial components~\cite{gmsys2023studying} for an accurate assessment. One such component can act as a controller, which is always active and responsible for the video streaming process's administration, security, and monitoring. Furthermore, traditional streaming approaches often result in the preparation and storage of rarely-requested segments~\cite{quortex2022mission}, contributing to accumulating useless segments on CDN cache servers.

\paragraph*{Regulation and standardization}
It is necessary to establish standardized guidelines for energy consumption in video streaming. Standards development organizations must ensure accessibility for infrastructure providers involved in physical facilities, equipment, and video streaming services. Additionally, standards facilitate the introduction of future regulations to reduce carbon emissions that may require infrastructure providers to adopt sustainable energy sources proportionate to their consumption for operating services. Consequently, the carbon efficiency of an infrastructure provider will become a crucial factor for adhering to the regulations~\cite{gmsys2023studying}.

% \paragraph*{Traced metrics for energy logging} 
% Developing traced metrics for logging energy usage can be beneficial to better comprehend the energy consumption associated with different components in the end-to-end video streaming process~\cite{gmsys2023studying}. However, it is essential to note that the act of logging itself can have an impact on energy consumption. Thus, it is crucial to consider a more efficient approach for determining when to log and when to skip to achieve a net reduction in carbon emissions. 

% \paragraph*{6G technology} In 6G technology,  a key design target for 6G includes zero energy devices and energy-optimized networks. There are key value indicators (KVIs) that are used to measure the impact of various aspects beyond just the basic performance indicators. For instance, sustainability is one of the dimensions that is evaluated through KVIs. One example of a KVI is Kleinrock's power metric that combines the concepts of 'goodness' (QoE) and 'badness' (\ch{CO2} emissions or energy consumption). However, in reality, it is difficult to accurately measure QoE and \ch{CO2} emissions~\cite{hossfeld2023greener}.

\paragraph*{Video prioritization} To mitigate carbon emissions, it is crucial to determine the paramount videos (\eg surveillance systems) and their required streaming times while limiting other trivial daily activities (\eg entertainment) to a reasonable duration, leading to less carbon emissions~\cite{tsp2019}.

\paragraph*{ML in video streaming} As ML gains prominence in estimating or mitigating the video streaming energy consumption~\cite{amirpour_optimizing_2023,liu2019energy,pandey2022energy}, there is a research gap in measuring the energy consumption during its training and testing phases.

\paragraph*{Environmental impact awareness of video streaming among users}
Gnanasekaran~\etal \cite{gnanasekaran2021digital} showed that most users are rarely aware of the environmental impact of digital services. Hence, making them aware is necessary to adopt a ``greener'' posture, for example, by selecting eco-friendly streaming and quality options considering each representation's carbon footprint. They can choose content with a lower carbon footprint, even if it compromises QoE to an acceptable degree. 
%An excellent example of this approach is the Carbonalyser plugin, which is based on the Shift project~\cite{ShiftProjectReport2020}. 

\paragraph*{Green service level agreement}
Inspired from~\cite{hossfeld2023greener}, one can apply ``carbon credits'' and measure the total \ch{CO2} emissions over a specific period (\eg month). Maintaining the trade-off between QoE and sustainability, users can purchase additional credits if they exceed their assigned limit, similar to buying extra data traffic volume. As suggested in the same article, service providers can offer users a combined service or experience level agreement that considers both carbon credits and expenses.

% ~\cite{patterson2021carbon}

\section{Concluding Remarks}
\label{sec:Conc}
The urgency of the climate crisis has increased the need to investigate the environmental impact of video streaming, a rapidly growing and popular digital activity. This survey evaluates the energy consumption and environmental impact of video content provisioning and consumption, aiming to understand the challenges and highlight opportunities to increase efficiency and reduce the damaging GHG emissions. We conducted a systematic literature review of \num{889} articles from relevant scientific sources using well-defined keywords and selection criteria. We selected \num{56} articles and approximately \num{200} references screened for relevance and clarity. We organized the surveyed articles into a taxonomy based on the most impactful components of the video streaming process on energy consumption: encoding, storing, retrieving, decoding, and displaying, and analyzed the state-of-the-art methods for optimizing their energy consumption. We discussed research gaps and open issues that need further research for a more energy-efficient and climate-friendly video streaming. The three most significant open issues include \begin{enumerate*} \item fixed bitrate ladders in HTTP live streaming, \item inefficient hardware utilization of existing video players, and \item lack of a reproducible energy measurement dataset covering various device types and coding parameters \end{enumerate*}. We believe that this survey offers valuable information for video researchers, engineers, and streaming service providers involved in the video streaming ecosystem.

%\section{Which Components Have a Greater Impact on Energy Consumption?}
% In this survey, after carefully investigating a diverse range of research papers focusing on energy consumption in video streaming, a crucial question emerges: "Which components have a greater impact on energy consumption?" By understanding the relative importance of different components, we can gain valuable insights into optimizing energy efficiency and devising more sustainable video streaming solutions. 

% The complexity of video on the end-user device has a negligible impact on power consumption, especially when compared to influential factors like brightness and contrast. Interestingly, even the size and power efficiency of the TV panel play a more significant role in determining power usage than the complexity of the scenes being viewed. This finding suggests that optimizing content solely for power consumption on playback devices may yield little benefit. Instead, the focus should be on optimizing content during encoding and distribution processes, as these areas offer more significant potential for achieving energy efficiency improvements.

% The study presented in~\cite{seeliger2022green} reveals that 

\bibliographystyle{ACM-Reference-Format}
\bibliography{ref%,sections/ref_hadi
}

\printnomenclature

\end{document}